\documentclass[reprint,superscriptaddress,amsmath,amssymb,aps,prl,floatfix]{revtex4-1}

\usepackage{hyperref}

\hypersetup{
  colorlinks=true,
  linkcolor=blue,
  citecolor=blue,
  urlcolor=blue
}

\newcommand{\Nrad}{N_{\mathrm{rad}}}
\newcommand{\pT}{\mathbf{p}_{\mathrm{T}}}
\newcommand{\pV}{\mathbf{p}_{\mathrm{V}}}
\newcommand{\pS}{\mathbf{p}_{\mathrm{S}}}
\newcommand{\Pvec}{\mathbf{P}}
\newcommand{\betag}{\boldsymbol{\beta}}
\newcommand{\epsV}{\epsilon_{\mathrm{V}}}
\newcommand{\epsS}{\epsilon_{\mathrm{S}}}

\begin{document}

\title{Eight Local Couplings of Gravitational Waves from Unified Field Equations}

\author{Hong-Bo Jin}
\email{hbjin@bao.ac.cn}
\affiliation{National Astronomical Observatories, Chinese Academy of Sciences, Beijing 100101, China}
\affiliation{The International Centre for Theoretical Physics Asia-Pacific, University of Chinese Academy of Sciences (UCAS), Beijing 100190, China}
\affiliation{Taiji Laboratory for Gravitational Wave Universe (Beijing/Hangzhou), UCAS, Beijing 100049, China}

\author{Yue-Liang Wu}
\email{ylwu@itp.ac.cn}
\affiliation{The International Centre for Theoretical Physics Asia-Pacific, University of Chinese Academy of Sciences (UCAS), Beijing 100190, China}
\affiliation{Taiji Laboratory for Gravitational Wave Universe (Beijing/Hangzhou), UCAS, Beijing 100049, China}
\affiliation{Institute of Theoretical Physics, Chinese Academy of Sciences, Beijing 100190, China}
\date{\today}
\begin{abstract}
A gravitational-wave (GW) detector records a linear mixture of local couplings under the assumption that extra polarizations enter geodesic deviation. Vacuum general relativity (GR) admits two transverse-traceless (TT) amplitudes.
It remains to determine the largest set of couplings that can sit in that mixture, and whether a tensor-only arm-length result selects GR uniquely.
The little group \(E(2)\) of a null four-momentum classifies the strain amplitudes \(\mathbf{p}=(p_{+},p_{\times},p_{x},p_{y},p_{b},p_{\ell})\), commonly written \(h_{P}\), which determine the electric tidal tensor along a ray. Geodesic deviation, recorded as differential arm length, therefore contains only those \(p_{P}\) that enter that tensor.
Lorentz mixing at helicity \(\pm1\) supplies a gravito-magnetic (GEM) field that does not enter the tidal tensor. if GEM field is static, that does not propagate as a wave. A time-varying helicity-\(\pm1\) current sources a GEM wave that enters the mixture as a coupling, to be isolated by its measured quantity.
For a radiation-zone wave that depends only on retarded time, \(\betag_{\perp}=\hat{\mathbf{k}}\times\partial_{t}(p_{x},p_{y})\), and the eight couplings are \(\Pvec=(\mathbf{p},\betag_{\perp})\).
Here we adopt unified field equations on \(\mathbf{p}\) to clarify the origin of each component of \(\Pvec\); the measured quantity of each coupling then isolates the polarizations in that mixture, which favors identification of distinct polarizations and model tests.
\end{abstract}

\maketitle

\section{Introduction}
\label{sec:intro}
\enlargethispage{2\baselineskip}

Direct detection of gravitational waves (GWs)~\cite{LIGO:2016} has made the polarization content of the detector response a measurable object.
A single interferometer, or a time-delay interferometry (TDI) combination, records a linear mixture of local couplings through its antenna patterns.
Networks isolate tensor, vector, and scalar channels in a stochastic background~\cite{Nishizawa:2009} and invert a compact-binary or continuous-wave mixture into those channels~\cite{Hayama:2013,Isi:2017,Takeda:2018}, under the assumption that every extra polarization enters geodesic deviation.
Vacuum general relativity (GR) admits two transverse-traceless (TT) amplitudes~\cite{Maggiore:2007}.
It remains to determine the largest set of local linear couplings that can sit in the mixture, and whether a tensor-only arm-length result selects GR uniquely.
We work in the asymptotically Minkowski radiation zone, in a local inertial frame. The retarded time \(u=t-z\) and \(\Box=\partial_{t}^{2}-\nabla^{2}\) refer to this background.

The strain amplitudes \(\mathbf{p}\) of helicities \(\pm2\), \(\pm1\), and \(0\) are classified by the little group \(E(2)\) of a null four-momentum, Eq.~\eqref{eq:E2}: a massless wave is classified by the stabilizer of \(k^{\mu}\) in \(\mathrm{SO}(1,3)\), not by a metric perturbation~\cite{Eardley:1973,Eardley:1973PRD}.
Geodesic deviation records the electric tidal tensor \(\mathcal{E}_{ij}\), which is linear in \(\partial_{u}^{2}\mathbf{p}\) along a ray.
Differential arm length records those \(p_{P}\) that enter \(\mathcal{E}_{ij}\).
Lorentz boosts mix this electric tidal response with a magnetic one.
A Fermi-frame rewriting of a GR plane wave remains helicity \(\pm2\) and is not an additional strain amplitude~\cite{Ruggiero:2020GEM,Ruggiero:2020res}.
Only helicity \(\pm1\) supplies a magnetic coupling that is not an entry of \(\mathcal{E}_{ij}\).
That coupling is the gravito-magnetic (GEM) field \(\betag_{\perp}=(\beta_{x},\beta_{y})\). A time-varying mass current or a radiating spin current sources the wave; a Sagnac loop records it as an optical-path difference, not by a direct arm-length readout. A laboratory test mass of velocity \(\mathbf{v}\) (or a laboratory spin) couples to the same field through the Mashhoon force~\cite{Mashhoon:2001}.
The eight couplings form
\begin{equation}
  \Pvec
  =
  \bigl(p_{+},\,p_{\times},\,p_{x},\,p_{y},\,p_{b},\,p_{\ell},\,\beta_{x},\,\beta_{y}\bigr)
  \label{eq:P8}
\end{equation}
The labels are plus, cross, vector-\(x\), vector-\(y\), breathing, longitudinal, and GEM. The strain entries are those commonly written \(h_{P}\).
The vector pair is \(\pV=(p_{x},p_{y})\) and \(\hat{\mathbf{k}}\) is the unit wave vector.
On a wave depending only on retarded time, \(\betag_{\perp}\) is given by Eq.~\eqref{eq:unify}.
The eight couplings are not eight independent degrees of freedom: \(\betag_{\perp}\) belongs to the same vector sector as \(\pV\). The number of outgoing strain branches is \(\Nrad\) in Table~\ref{tab:Nrad}~\cite{Brans:1961,Jacobson:2004,Will:2014,Dong:2024,Gao:2025GW}.
When the vector sector propagates, the mixture can include a GEM wave absent from the arm-length channel.
The field equations~\eqref{eq:eom} govern \(\mathbf{p}\) in the Minkowski radiation zone, onto which a numerical-relativity or post-Newtonian waveform, and a high-frequency calculation on a curved background~\cite{Isaacson:1968}, match at large distance.

\section{Polarization content from the little group}
\label{sec:eight}

\subsection{Tidal amplitudes and GEM}
\label{sec:obs}

The Lorentz group \(\mathrm{SO}(1,3)\) acts on Minkowski spacetime.
A radiation-zone GW is massless: the four-momentum \(k^{\mu}\) obeys \(k_{\mu}k^{\mu}=0\) with \(k^{\mu}\neq0\).
Greek indices run over \(0,1,2,3\); Latin indices over \(x,y,z\).
A superscript on a four-vector is a Lorentz index, so \(k^{2}\) is the \(y\) component of \(k^{\mu}\); the Minkowski square is \(k_{\mu}k^{\mu}\).
The speed of light is restored in phenomenological formulae and set to \(c=1\) in the radiation-zone algebra below.
Coordinates are Cartesian; \(\nabla^{2}\) is the flat Laplacian.
The little group of \(k^{\mu}\) is the stabilizer
\begin{equation}
\begin{gathered}
  E(2)
  =
  G_{k}
  =
  \bigl\{\Lambda\in\mathrm{SO}(1,3)\mid \Lambda k=k\bigr\},
  \\
  k_{\mu}k^{\mu}=0,
  \quad
  k^{\mu}\neq0.
\end{gathered}
  \label{eq:E2}
\end{equation}
which is isomorphic to the two-dimensional Euclidean group~\cite{Eardley:1973,Eardley:1973PRD}.
The flag \(\tau_{P}\) equals \(1\) if the strain entry \(p_{P}\) enters \(\mathcal{E}_{ij}\) and equals \(0\) otherwise, with \(P\in\{+,{\times},x,y,b,\ell\}\).

\begin{table}[!htb]
\caption{\label{tab:Nrad}
Radiation content of \(\Pvec\). \(\Nrad\) is the number of outgoing strain branches. The interferometer (IFO) column lists the amplitudes with \(\tau_{P}=1\). GEM is present whenever \(\epsV=1\), independently of \(\tau_{x},\tau_{y}\). If \(\pV\) is present, \(\betag_{\perp}\) follows from~\eqref{eq:unify} on a wave depending only on retarded time.}
\begin{ruledtabular}
\begin{tabular}{lcccc}
Theory & \((\epsV,\epsS)\) & \(\Nrad\) & IFO & GEM \\
GR & \((0,0)\) & \(2\) & \(p_{+},p_{\times}\) & \(0\) \\
scalar-tensor & \((0,1)\) & \(3\) & \(\pT\), scalar & \(0\) \\
Einstein--{\AE}ther\footnotemark[1] & \((1,1)\) & \(5\) & \(\pT,\pV\), scalar & \(\betag_{\perp}\) \\
GQFT\footnotemark[2] & \((1,1)\) & \(5\) & \(\pT\), \(p_{b}\) & \(\betag_{\perp}\) \\
maximal\footnotemark[3] & \((1,1)\) & \(\le 6\) & all \(p_{P}\) & \(\betag_{\perp}\) \\
\end{tabular}
\end{ruledtabular}
\footnotetext[1]{One independent scalar mode; \(\tau_{x}=\tau_{y}=1\)~\cite{Jacobson:2004}.}
\footnotetext[2]{Gravitational quantum field theory (GQFT): \(\tau_{x}=\tau_{y}=0\)~\cite{Gao:2025GW}.}
\footnotetext[3]{IFO lists every \(p_{P}\) with \(\tau_{P}=1\).}
\end{table}

The classification applies to each massless helicity on its null cone.
Along \(+\hat{\mathbf{z}}\) at speed \(c\), \(k^{\mu}=\omega(1,0,0,1)\) with angular frequency \(\omega\neq0\), so \(k^{1}=k^{2}=0\) while \(k^{\mu}\neq0\) and \(k_{\mu}k^{\mu}=0\).
Helicity \(\pm2\) gives the tensor pair, \(\pm1\) the vector pair, and \(0\) the two scalars (breathing and longitudinal).
The GEM pair of a wave along \(\hat{\mathbf{z}}\) is \(\betag_{\perp}=(\beta_{x},\beta_{y})\), orthogonal to the propagation direction, with no independent \(\beta_{z}\).
In a local Lorentz frame the geodesic-deviation equation for a pair of test masses with spatial separation \(\xi^{i}\)
\begin{equation}
  \delta\ddot\xi^{i}
  =
  -\mathcal{E}^{i}{}_{j}\,\xi^{j}
  \label{eq:geo}
\end{equation}
measures a symmetric tidal tensor \(\mathcal{E}_{ij}\) with six independent components; \(\betag_{\perp}\) is not an entry of \(\mathcal{E}_{ij}\), so geodesic deviation does not couple to the GEM force.
The amplitudes
\begin{equation}
  \mathbf{p}
  =
  \bigl(
  \underbrace{p_{+},\,p_{\times}}_{\pT},\;
  \underbrace{p_{x},\,p_{y}}_{\pV},\;
  \underbrace{p_{b},\,p_{\ell}}_{\pS}
  \bigr)
  \label{eq:pdef}
\end{equation}
are the Eardley strain polarizations, commonly denoted \(h_{P}\).
Along a wave depending only on retarded time they determine \(\mathcal{E}_{ij}\) by second retarded derivatives: \(\mathcal{E}_{xx}-\mathcal{E}_{yy}\), \(\mathcal{E}_{xy}\), \(\mathcal{E}_{xz}\), \(\mathcal{E}_{yz}\), \(\mathcal{E}_{xx}+\mathcal{E}_{yy}\), and \(\mathcal{E}_{zz}\) are linear in \(\partial_{u}^{2}p_{+}\), \(\partial_{u}^{2}p_{\times}\), \(\partial_{u}^{2}p_{x}\), \(\partial_{u}^{2}p_{y}\), \(\partial_{u}^{2}p_{b}\), and \(\partial_{u}^{2}p_{\ell}\), respectively.
When \(\mathcal{E}_{xz}\) and \(\mathcal{E}_{yz}\) are present, geodesic deviation therefore records \(\partial_{u}^{2}\pV\), equivalently \(\partial_{u}\betag_{\perp}\) by~\eqref{eq:unify}, not the GEM force itself.
\(\pT\), \(\pV\), and \(\pS\) are the tensor, vector, and scalar pairs.
Tensor indices run over \(A\in\{+,{\times}\}\), vector indices over \(I\in\{x,y\}\), scalar indices over \(\alpha\in\{b,\ell\}\), and the collective index \(P\in\{+,{\times},x,y,b,\ell\}\).
A rotation by an angle \(\vartheta\) about \(\hat{\mathbf{z}}\) acts as \(p_{+}+ip_{\times}\to e^{2i\vartheta}(p_{+}+ip_{\times})\), \(p_{x}+ip_{y}\to e^{i\vartheta}(p_{x}+ip_{y})\), and leaves \((p_{b},p_{\ell})\) invariant.
The flag \(\tau_{P}\) selects which of these amplitudes enter \(\mathcal{E}_{ij}\).

A laboratory test mass of velocity \(v_{j}\), or a laboratory spin, couples to a gravito-magnetic field \(\betag\) through the force \(F^{i}=m\,\varepsilon^{ijk}v_{j}\beta_{k}\) (or a precession \(\boldsymbol{\Omega}\propto\betag\)), with \(\varepsilon_{ijk}\) the Levi-Civita symbol, \(\varepsilon_{xyz}=+1\); this \(F^{i}\) is distinct from the interferometer antenna patterns \(F_{P}\)~\cite{Mashhoon:2001}.
A Lorentz boost mixes the electric tidal tensor \(\mathcal{E}_{ij}\) with this magnetic response, as a Lorentz boost mixes \(\mathbf{E}\) and \(\mathbf{B}\).
Helicity \(\pm1\) therefore has a strain pair \(\pV\) and a magnetic pair \(\betag_{\perp}\).
The magnetic tidal field of helicity \(\pm2\) is the Fermi-frame rewriting of \(\pT\), not an additional component of \(\Pvec\).
A source for the helicity-\(\pm1\) pair is a time-varying transverse current: Thorne's \(\dot J\)~\cite{Thorne:1980}, or, in a theory that radiates net fermion spin, \(\dot{\mathbf{U}}\)~\cite{Gao:2025GW}; in the center-of-mass frame the conserved total momentum does not radiate, Eq.~\eqref{eq:Jdef}.

\subsection{Relation of GEM to \(\pV\)}
\label{sec:faraday}

On a radiation-zone wave depending only on retarded time \(u=t-z\),
\begin{equation}
  \betag_{\perp}
  =
  \hat{\mathbf{k}}\times\partial_{u}\pV,
  \label{eq:unify}
\end{equation}
with \(\hat{\mathbf{k}}=\hat{\mathbf{z}}\) the unit wave vector, so \(\beta_{x}=-\partial_{u}p_{y}\) and \(\beta_{y}=\partial_{u}p_{x}\).
This is the analog of \(\mathbf{B}=\nabla\times\mathbf{A}\) with vector strain \(\pV\) in the role of the transverse vector potential.
Both entries have helicity \(\pm1\) and belong to the same vector sector.
When \(\mathcal{E}_{xz}\) and \(\mathcal{E}_{yz}\) are sourced, differential arm length records \(\pV(u)\) and geodesic deviation records \(\mathcal{E}_{ij}\propto\partial_{u}^{2}\pV\). A Sagnac loop records \(\partial_{u}\pV\) (equivalently \(\betag_{\perp}\)) whether or not those electric components are present.
The GEM force itself is invisible to a rest-mass Michelson combination in every model that has \(\betag_{\perp}\). What varies by theory is whether the same helicity-\(\pm1\) current also appears in \(\mathcal{E}_{ij}\) through \(\pV\).
A model may set \(\tau_{x}=\tau_{y}=0\) even if \(\pV\) propagates: rest-mass Michelson combinations then read \(\mathcal{E}_{ij}\sim R_{i00j}=0\) in the vector sector, Table~\ref{tab:Nrad}.
Einstein--{\AE}ther and GQFT share \(\Nrad=5\) in that table and differ in this assignment (\(\tau_{x}=\tau_{y}=1\) versus \(0\))~\cite{Jacobson:2004,Gao:2025GW}.
The GQFT row is an existence example that a GEM wave can sit in the mixture without a geodesic-deviation vector; no particular model is tested here.
Appendix~\ref{sec:gqft} evaluates that row, including the near-zone Lense--Thirring piece excluded from radiative \(\Pvec\) and the outgoing \(\dot{\mathbf{U}}\) branch.

For a monochromatic plane wave (convention \(e^{-i\omega u}\), hats denoting Fourier amplitudes),
\begin{equation}
  \hat\beta_{x}=i\omega\,\hat p_{y},
  \qquad
  \hat\beta_{y}=-i\omega\,\hat p_{x}.
  \label{eq:betaPhase}
\end{equation}
The factor \(i=e^{i\pi/2}\) is a phase shift of \(\pi/2\); the squared amplitudes of \(\pV\) and \(\betag_{\perp}\) alternate, as in the plane-wave form of Appendix~\ref{sec:plane}.
The map is invertible for \(\omega\neq0\).
On a broadband waveform, \(\partial_{u}\pV\) is the local rate of change of \(\pV\).

\section{Unified field equations}
\label{sec:eom}

\subsection{Unified operator}
\label{sec:L}

A helicity-preserving linear operator on \(\mathbf{p}\) is block diagonal on the tensor, vector, and scalar pairs \(\{\mathrm{T},\mathrm{V},\mathrm{S}\}\).
Each pair either propagates or is a constraint.
The switches \(\epsV,\epsS\in\{0,1\}\) select a wave operator (\(\epsilon=1\)) or a constraint (\(\epsilon=0\)); these \(\epsilon\) parameters are distinct from the Levi-Civita symbol \(\varepsilon_{ijk}\).
On each pair \(\sigma\in\{\mathrm{T},\mathrm{V},\mathrm{S}\}\), a value of \(\epsilon_{\sigma}\) other than \(0\) or \(1\) would introduce a Klein--Gordon mass and is omitted.
The operator is
\begin{equation}
  L_{\sigma}
  \;=\;
  \epsilon_{\sigma}\,\Box_{\sigma}
  +(1-\epsilon_{\sigma})\,\mu_{\sigma}^{2},
  \qquad
  \Box_{\sigma}
  \;=\;
  \partial_{t}^{2}-c_{\sigma}^{2}\nabla^{2},
  \label{eq:Lsig}
\end{equation}
with \(\epsilon_{\mathrm{T}}\equiv1\), \(c_{\mathrm{T}}=1\) the tensor propagation speed in units of \(c\), as constrained by the binary neutron-star event GW170817~\cite{LIGO:2017GW170817}.
Here \(c_{\sigma}\) is the speed of sector \(\sigma\), distinct from the inertial mass \(m\) in the GEM force, and \(\mu_{\sigma}^{2}>0\) is a constraint parameter.
A propagating pair is massless on the null cone of \(\Box_{\sigma}\); if \(c_{\sigma}\neq1\) that cone is not the Minkowski light cone of~\eqref{eq:E2}.
The source coupling of sector \(\sigma\) is \(\lambda_{\sigma}\), and \(J_{\sigma}\) are the corresponding projections of the stress tensor \(T^{\mu\nu}\), given in~\eqref{eq:Jdef}.
The field equation on \(\mathbf{p}\) is
\begin{equation}
  \boldsymbol{\mathcal{L}}(\epsV,\epsS)\,\mathbf{p}
  =
  \mathbf{J}_{\mathbf{p}}(\epsV,\epsS),
  \label{eq:eom}
\end{equation}
with \(\boldsymbol{\mathcal{L}}=\mathrm{diag}(\Box,\Box,L_{\mathrm{V}},L_{\mathrm{V}},L_{\mathrm{S}},L_{\mathrm{S}})\), where \(\Box\equiv\Box_{\mathrm{T}}\) is the tensor d'Alembertian.
In components,
\begin{equation}
\begin{aligned}
  \Box\,p_{A}
  &=
  \lambda_{\mathrm{T}}\,J_{A},
  \\
  \bigl[\epsV\Box_{\mathrm{V}}+(1-\epsV)\mu_{\mathrm{V}}^{2}\bigr]\,p_{I}
  &=
  \epsV\,\lambda_{\mathrm{V}}\,J_{I},
  \\
  \bigl[\epsS\Box_{\mathrm{S}}+(1-\epsS)\mu_{\mathrm{S}}^{2}\bigr]\,p_{\alpha}
  &=
  \epsS\,\lambda_{\mathrm{S}}\,J_{\alpha}.
\end{aligned}
\label{eq:eomcomp}
\end{equation}
On a wave depending only on \(u\), the GEM pair is fixed by~\eqref{eq:unify}, or equivalently \(\partial_{u}\pV+\hat{\mathbf{k}}\times\betag_{\perp}=0\), and follows from \(\pV\) rather than entering~\eqref{eq:eom} as an additional unknown.
If a realization mixes the two scalar branches, the scalar line is replaced by a \(2\times2\) matrix \(M\) acting on \((p_{b},p_{\ell})\).

Sources sit in the same little-group representation, including current-multipole (\(\dot J\)) contributions~\cite{Thorne:1980}:
\begin{equation}
\begin{aligned}
  J_{A}
  &=
  \Pi_{A}^{ij}
  \bigl(
  \ddot Q_{ij}
  +\mathcal{M}_{ij}(\dot J,\ddot J)
  \bigr),
  \\
  J_{I}
  &=
  \Pi_{\perp I}{}^{j}\,S^{j}[\dot J],
  \\
  J_{\alpha}
  &=
  \Pi_{\alpha}\,T.
\end{aligned}
\label{eq:Jdef}
\end{equation}
Here \(Q_{ij}\) is the mass quadrupole (isolated sources have no mass dipole), \(\mathcal{M}_{ij}\) collects mass-current multipoles, \(S^{j}[\dot J]\) is the leading current-multipole source in the center-of-mass frame, \(T=\eta_{\mu\nu}T^{\mu\nu}\) is the trace, and \(\Pi_{A}^{ij}\), \(\Pi_{\perp I}{}^{j}\), and \(\Pi_{\alpha}\) are the Eardley projectors onto a wave along \(\hat{\mathbf{k}}\)~\cite{Eardley:1973,Eardley:1973PRD}.
The densities \(J_{A}\), \(J_{I}\), \(J_{\alpha}\) are not the angular momentum whose derivative is written \(\dot J\).
The current \(S^{j}[\dot J]\) is a time-varying transverse source for the helicity-\(\pm1\) pair~\cite{Thorne:1980,Gao:2025GW}.
When \(\epsV=\epsS=0\), the retarded tensor solution reproduces the Einstein multipole formula~\cite{Thorne:1980,Maggiore:2007},
\begin{equation}
  p_{A}
  \;=\;
  \frac{2G}{c^{4}r}\,
  \Pi_{A}^{ij}
  \bigl(
  \ddot Q_{ij}
  +\mathcal{M}_{ij}(\dot J,\ddot J)
  \bigr),
  \label{eq:multipole}
\end{equation}
with \(G\) Newton's constant and \(r\) the distance to the source.
When \(\epsV=1\) a helicity-\(\pm1\) current feeds \(\pV\) and, by~\eqref{eq:unify} on a wave depending only on retarded time, \(\betag_{\perp}\).
Appendix~\ref{sec:op} derives Eqs.~\eqref{eq:eomcomp} and~\eqref{eq:Jdef}; Appendix~\ref{sec:green} records the retarded kernel of each entry of \(\Pvec\); Appendix~\ref{sec:helmS} records the Helmholtz and paraxial reductions of each propagating pair; Appendix~\ref{sec:dict} records a dictionary that expands Table~\ref{tab:Nrad}, including the GQFT existence example.

After the near-zone motion has been reduced to \(\ddot Q_{ij}\) and \(\dot J\), the mass quadrupole is the source of \(p_{+}\) and \(p_{\times}\)~\cite{Thorne:1980,Maggiore:2007}.
If \(\epsV=1\), a helicity-\(\pm1\) current also feeds \(\pV\) and \(\betag_{\perp}\).
A vanishing stress-tensor trace leaves \(\pS\) unexcited unless an additional scalar source is present.

\subsection{Measured time series}
\label{sec:meas}

The field equations determine the radiation-zone amplitudes \(p_{P}(u)\). \(\betag_{\perp}\) follows from~\eqref{eq:unify}.
The recorded quantities are the linear instrument responses of those amplitudes, and are the measured quantities of each coupling.
A Michelson or TDI combination records a differential arm length \(\Delta L\), equivalently an interferometer strain \(s_{\mathrm{IFO}}\), from the Eardley antenna patterns \(F_{P}\)~\cite{Eardley:1973,Eardley:1973PRD}, with \(\hat{\mathbf{n}}\) the source direction on the sky.
A GEM readout records a projection of \(\betag_{\perp}\), as the inertial-sensor acceleration of a test mass with velocity \(\mathbf{v}\), as the optical-path difference of a Sagnac loop of vector area \(\mathbf{A}\), or as a laboratory-magnet precession \(\boldsymbol{\Omega}\propto\betag\),
\begin{equation}
\begin{aligned}
  s_{\mathrm{IFO}}(t)
  &=
  \sum_{P:\,\tau_{P}=1}F_{P}(\hat{\mathbf{n}})\,p_{P}(t),
  \\
  \mathbf{a}
  &=
  \mathbf{v}\times\betag_{\perp},
  \qquad
  \delta\ell
  =
  \frac{2\,\mathbf{A}\cdot\betag_{\perp}}{c}.
\end{aligned}
\label{eq:meas}
\end{equation}
The antenna patterns \(F_{P}\) are distinct from the GEM force \(F^{i}\).
A search is a matched filter of these templates against the data, using the same source hypothesis that produced \(\mathbf{p}(u)\).
The time derivative of \(\Delta L\), or a factor \(i\omega\) applied to a tidal spectrum, is a rewriting of the arm-length response and is not a GEM measurement.
Appendix~\ref{sec:stemp} records the GEM time series of~\eqref{eq:meas} and Appendix~\ref{sec:plane} the plane-wave solutions. \(\betag_{\perp}\) is then determined by \(\pV\).

\section{Discussion and conclusions}
\label{sec:concl}

A detector records a linear mixture of local couplings to a radiation-zone wave.
The little group \(E(2)\) supplies the Eardley strain amplitudes \(\mathbf{p}\). Lorentz mixing at helicity \(\pm1\) supplies \(\betag_{\perp}\).
Differential arm length records \(s_{\mathrm{IFO}}\) in~\eqref{eq:meas}, built only from the entries with \(\tau_{P}=1\). A Sagnac loop records \(\delta\ell\); an inertial sensor records \(\mathbf{a}\); a laboratory magnet records \(\boldsymbol{\Omega}\).
For a radiation-zone wave that depends only on retarded time, \(\betag_{\perp}=\hat{\mathbf{k}}\times\partial_{t}(p_{x},p_{y})\), and the eight couplings are \(\Pvec=(\mathbf{p},\betag_{\perp})\).

Here we adopt unified field equations on \(\mathbf{p}\) that assign the tensor pair to a mass quadrupole, the vector and GEM entries to a helicity-\(\pm1\) current, and the scalar pair to the stress-tensor trace. The measured quantity of each coupling then isolates those components in the mixture, which favors identification of distinct polarizations and model tests. A tensor-only arm-length result therefore does not select GR uniquely: the mixture can still contain a GEM wave, sourced by a time-varying helicity-\(\pm1\) current and recorded by a Sagnac loop; it is also compatible with a propagating vector sector that does not enter geodesic deviation, the GQFT row of Table~\ref{tab:Nrad}.
Near-zone Lense--Thirring fields, as measured with laser-ranged satellites~\cite{Ciufolini:2000}, and radiative \(\betag_{\perp}\) belong to the same helicity-\(\pm1\) sector. GR constrains the outgoing branch, while a propagating vector sector lets that branch appear in the mixture. That radiative branch remains helicity \(\pm1\): it is the outgoing continuation of near-zone GEM, not an additional geodesic-deviation mode.

Pure-polarization tests~\cite{Abbott:2017GW170814,Takeda:2018,Takeda:2021} favor tensor modes over pure vector or scalar alternatives, restricting extreme assignments of \((\epsV,\epsS)\). Those analyses use only the arm-length channel and cannot replace identification of the polarizations in the mixture by the measured quantity of each coupling.
Any other linearized theory is read onto Table~\ref{tab:Nrad} by the dictionary of Appendix~\ref{sec:dict}, which expands that table.

\begin{acknowledgments}
  We thank colleagues in the gravitational-wave community for discussions.
  This work is funded by the National Astronomical Observatories of the Chinese Academy of Sciences, Project No.~E4TG6601, and has been supported in part by the National Key Research and Development Program of China under Grant No.~2021YFC2203000.
\end{acknowledgments}

\clearpage
\phantomsection
\label{app:sm}
\setcounter{secnumdepth}{3}
\setcounter{section}{0}
\setcounter{equation}{0}
\setcounter{table}{0}
\renewcommand{\thesection}{S\arabic{section}}
\renewcommand{\theequation}{S\arabic{equation}}
\renewcommand{\thetable}{S\arabic{table}}
\renewcommand{\thesubsection}{S\arabic{section}.\arabic{subsection}}

\section*{Supplemental Material}
\addcontentsline{toc}{section}{Supplemental Material}

\section{Scope}
\label{sec:scope}

No metric perturbation \(h_{\mu\nu}\) and no particular action are assumed.
A radiation-zone gravitational wave is a massless excitation in an asymptotically Minkowski local inertial frame~\cite{Eardley:1973,Eardley:1973PRD}.
Its independent labels are the local linear couplings assembled into
\begin{equation}
  \Pvec
  =
  \bigl(p_{+},\,p_{\times},\,p_{x},\,p_{y},\,p_{b},\,p_{\ell},\,\beta_{x},\,\beta_{y}\bigr),
  \label{eq:P8S}
\end{equation}
The labels are plus, cross, vector-\(x\), vector-\(y\), breathing, longitudinal, and the two gravito-magnetic (GEM) entries of Eq.~\eqref{eq:P8}.
The vector pair is \(\pV=(p_{x},p_{y})\).
A concrete theory is specified only later, by which of these labels propagate, which of them enter the electric tidal tensor \(\mathcal{E}_{ij}\), and which sources are present.
That assignment is Table~\ref{tab:Nrad}.
This appendix derives the equations that Table~\ref{tab:Nrad} specializes, records the solution of each of the eight couplings, and states how another model is read onto the same table.
The operator of Eqs.~\eqref{eq:Lsig} and~\eqref{eq:eomcomp}, the sources of Eq.~\eqref{eq:Jdef}, and the retarded kernel occupy the next three sections.
Helmholtz and paraxial propagation, the recorded time series of Eq.~\eqref{eq:meas}, and the plane-wave form of the eight couplings follow.
The gravitational quantum field theory (GQFT) row and the dictionary for Table~\ref{tab:Nrad} close this appendix.

\section{\texorpdfstring{Six strain amplitudes from \(E(2)\) and geodesic deviation}{Six strain amplitudes from E(2) and geodesic deviation}}
\label{sec:E2}

A radiation-zone four-momentum obeys \(k_{\mu}k^{\mu}=0\) with \(k^{\mu}\neq0\).
Its stabilizer in \(\mathrm{SO}(1,3)\) is the little group \(E(2)\), Eq.~\eqref{eq:E2}~\cite{Eardley:1973,Eardley:1973PRD}.
Irreducible representations on a null cone are labeled by helicity in \(\{0,\pm1,\pm2\}\) in four dimensions.
Helicity \(\pm2\) supplies a complex amplitude \(p_{+}+ip_{\times}\), helicity \(\pm1\) a complex amplitude \(p_{x}+ip_{y}\), and helicity \(0\) a pair \((p_{b},p_{\ell})\).
A rotation by \(\vartheta\) about the ray therefore acts as
\begin{equation}
  p_{+}+ip_{\times}\to e^{2i\vartheta}(p_{+}+ip_{\times}),
  \qquad
  p_{x}+ip_{y}\to e^{i\vartheta}(p_{x}+ip_{y}),
  \label{eq:rot}
\end{equation}
and leaves \((p_{b},p_{\ell})\) invariant.
These six real functions are the Eardley strain polarizations \(\mathbf{p}\), commonly written \(h_{P}\), with collective index \(P\in\{+,{\times},x,y,b,\ell\}\).
The tensor, vector, and scalar pairs are \(\pT=(p_{+},p_{\times})\), \(\pV=(p_{x},p_{y})\), and \(\pS=(p_{b},p_{\ell}\).
Tensor indices run over \(A\in\{+,{\times}\}\), vector indices over \(I\in\{x,y\}\), and scalar indices over \(\alpha\in\{b,\ell\}\).

Geodesic deviation of a pair of rest masses with separation \(\xi^{i}\),
\begin{equation}
  \delta\ddot\xi^{i}
  =
  -\mathcal{E}^{i}{}_{j}\,\xi^{j},
  \label{eq:geoS}
\end{equation}
measures the symmetric electric tidal tensor \(\mathcal{E}_{ij}\sim R_{i00j}\), which has six independent components.
On a wave along \(\hat{\mathbf{z}}\) depending only on \(u=t-z\), the Riemann symmetries reduce \(\mathcal{E}_{ij}\) to second retarded derivatives of \(\mathbf{p}\):
\begin{equation}
\begin{aligned}
  \mathcal{E}_{xx}-\mathcal{E}_{yy}
  &\propto
  \partial_{u}^{2}p_{+},
  &
  \mathcal{E}_{xy}
  &\propto
  \partial_{u}^{2}p_{\times},
  \\
  \mathcal{E}_{xz}
  &\propto
  \partial_{u}^{2}p_{x},
  &
  \mathcal{E}_{yz}
  &\propto
  \partial_{u}^{2}p_{y},
  \\
  \mathcal{E}_{xx}+\mathcal{E}_{yy}
  &\propto
  \partial_{u}^{2}p_{b},
  &
  \mathcal{E}_{zz}
  &\propto
  \partial_{u}^{2}p_{\ell}.
\end{aligned}
\label{eq:Emap}
\end{equation}
Equation~\eqref{eq:Emap} is a dictionary between the electric tidal tensor and the strain labels, not a field equation.
It holds whenever the corresponding Riemann components are generated by that \(p_{P}\).
The flag \(\tau_{P}\) equals \(1\) if \(p_{P}\) enters \(\mathcal{E}_{ij}\) and equals \(0\) otherwise.
Geodesic deviation, recorded as differential arm length, therefore contains only those \(p_{P}\) that enter that tensor,
\begin{equation}
  s_{\mathrm{IFO}}(t)
  =
  \sum_{P:\,\tau_{P}=1}F_{P}(\hat{\mathbf{n}})\,p_{P}(t),
  \label{eq:ifo}
\end{equation}
with \(F_{P}\) the Eardley antenna patterns and \(\hat{\mathbf{n}}\) the source direction on the sky~\cite{Eardley:1973,Eardley:1973PRD}.
Vacuum GR has \(\tau_{+}=\tau_{\times}=1\) and \(\tau_{P}=0\) otherwise.
A model may set \(\tau_{x}=\tau_{y}=0\) even if \(p_{x}\) and \(p_{y}\) propagate, which is the GQFT row of Table~\ref{tab:Nrad}.

\section{GEM from Lorentz mixing}
\label{sec:lorentz}

A Lorentz boost mixes \(\mathcal{E}_{ij}\) with a magnetic response, as it mixes \(\mathbf{E}\) and \(\mathbf{B}\).
For helicity \(\pm2\) that magnetic tidal field is a Fermi-frame rewriting of \(\pT\) and is not an extra label~\cite{Ruggiero:2020GEM,Ruggiero:2020res}.
For helicity \(\pm1\) a laboratory test mass of velocity \(v_{j}\), or a laboratory spin, couples through the force
\begin{equation}
  F^{i}
  =
  m\,\varepsilon^{ijk}v_{j}\beta_{k},
  \label{eq:Fgem}
\end{equation}
or a precession \(\boldsymbol{\Omega}\propto\betag\), with \(\varepsilon_{ijk}\) the Levi-Civita symbol, \(\varepsilon_{xyz}=+1\)~\cite{Mashhoon:2001}.
This pair \(\betag_{\perp}=(\beta_{x},\beta_{y})\), orthogonal to the ray, is not an entry of \(\mathcal{E}_{ij}\): geodesic deviation~\eqref{eq:geoS} is insensitive to \(v\times\betag\).
There is no independent \(\beta_{z}\) at this order, because a boost along the ray does not generate a new longitudinal magnetic coupling in the little-group decomposition.

On a wave that depends only on retarded time \(u=t-z\), transversality and the helicity-\(\pm1\) analog of \(\mathbf{B}=\hat{\mathbf{k}}\times\mathbf{E}\) lock the magnetic pair to one retarded derivative of the strain pair, with \(\hat{\mathbf{k}}\) the unit wave vector:
\begin{equation}
  \betag_{\perp}
  =
  \hat{\mathbf{k}}\times\partial_{u}\pV,
  \qquad
  \beta_{x}=-\partial_{u}p_{y},
  \qquad
  \beta_{y}=\partial_{u}p_{x}.
  \label{eq:unifyS}
\end{equation}
This is Eq.~\eqref{eq:unify}.
The relation is kinematics on the null cone of the vector pair.
The map is invertible at nonzero angular frequency \(\omega\), so \(\betag_{\perp}\) follows from \(\pV\) and does not add outgoing branches.
A Sagnac loop records \(\partial_{u}\pV\) (equivalently \(\betag_{\perp}\)) whether or not \(\tau_{x}=\tau_{y}=1\).
Quasi-static Lense--Thirring fields belong to the same helicity-\(\pm1\) sector in the static limit~\cite{Ciufolini:2000}; radiative \(\Pvec\) contains the outgoing \(1/r\) branch, recovered from~\eqref{eq:unifyS} when the vector pair propagates.

\section{Helicity-preserving operator: derivation of Eqs.~(7) and~(9)}
\label{sec:op}

A linear homogeneous operator on \(\mathbf{p}\) that commutes with~\eqref{eq:rot} cannot mix distinct helicities.
It is therefore block diagonal on the tensor, vector, and scalar pairs \(\{\mathrm{T},\mathrm{V},\mathrm{S}\}\).
Each pair is acted on by a scalar operator equally on the two real components of that helicity (or on each scalar branch).
On each pair \(\sigma\in\{\mathrm{T},\mathrm{V},\mathrm{S}\}\), the unique second-order, Lorentz-invariant, massless operator on that cone is the d'Alembertian
\begin{equation}
  \Box_{\sigma}
  =
  \partial_{t}^{2}-c_{\sigma}^{2}\nabla^{2},
  \label{eq:box}
\end{equation}
with \(c_{\sigma}\) the propagation speed of pair \(\sigma\).
A Klein--Gordon mass on a propagating pair would replace \(E(2)\) by \(\mathrm{SO}(3)\) and is omitted~\cite{Eardley:1973,Eardley:1973PRD}.
The unique alternative of the same order that removes the pair from the radiation zone is an algebraic constraint \(\mu_{\sigma}^{2}p_{\sigma}=0\) with \(\mu_{\sigma}^{2}>0\).
The interpolation that realizes either option, and not a massive pole, is
\begin{equation}
  L_{\sigma}
  =
  \epsilon_{\sigma}\,\Box_{\sigma}
  +(1-\epsilon_{\sigma})\,\mu_{\sigma}^{2},
  \qquad
  \epsilon_{\sigma}\in\{0,1\},
  \label{eq:LsigS}
\end{equation}
which is Eq.~\eqref{eq:Lsig}, with \(\epsilon_{\mathrm{T}}\equiv1\) and \(c_{\mathrm{T}}=1\)~\cite{LIGO:2017GW170817}.
A value of \(\epsilon_{\sigma}\) other than \(0\) or \(1\) yields a massive wave operator and is excluded by the same little-group requirement.
If a realization mixes the two scalar branches, \(L_{\mathrm{S}}\) is replaced by a \(2\times2\) matrix \(M\) on \((p_{b},p_{\ell})\).

Coupling each pair to a source in the same representation, with source coupling \(\lambda_{\sigma}\) and source density \(\mathbf{J}_{\sigma}\), gives
\begin{equation}
  L_{\sigma}\,\mathbf{p}_{\sigma}
  =
  \epsilon_{\sigma}\,\lambda_{\sigma}\,\mathbf{J}_{\sigma},
  \label{eq:eomblock}
\end{equation}
or, in components, Eq.~\eqref{eq:eomcomp}:
\begin{equation}
\begin{aligned}
  \Box\,p_{A}
  &=
  \lambda_{\mathrm{T}}\,J_{A},
  \\
  \bigl[\epsV\Box_{\mathrm{V}}+(1-\epsV)\mu_{\mathrm{V}}^{2}\bigr]\,p_{I}
  &=
  \epsV\,\lambda_{\mathrm{V}}\,J_{I},
  \\
  \bigl[\epsS\Box_{\mathrm{S}}+(1-\epsS)\mu_{\mathrm{S}}^{2}\bigr]\,p_{\alpha}
  &=
  \epsS\,\lambda_{\mathrm{S}}\,J_{\alpha}.
\end{aligned}
\label{eq:eomS}
\end{equation}
The GEM pair is recovered from \(\pV\) by~\eqref{eq:unifyS} and does not enter~\eqref{eq:eomS} as an additional unknown.
The number of independent outgoing strain branches is therefore
\begin{equation}
  \Nrad(\epsV,\epsS)
  =
  2+2\,\Theta(\epsV)+n_{\mathrm{S}}(\epsS),
  \label{eq:NradS}
\end{equation}
where \(\Theta(\epsilon)=1\) for \(\epsilon=1\) and \(0\) for \(\epsilon=0\), \(n_{\mathrm{S}}(0)=0\), and \(n_{\mathrm{S}}(1)=1\) or \(2\).
This count, not the eight entries of \(\Pvec\), is what Table~\ref{tab:Nrad} reports.

\section{Sources in the same representation: derivation of Eq.~(10)}
\label{sec:src}

The stress tensor \(T^{\mu\nu}\) decomposes under \(E(2)\) into the same helicities as \(\mathbf{p}\)~\cite{Thorne:1980,Eardley:1973PRD}.
Helicity \(\pm2\) is sourced by the mass quadrupole \(Q_{ij}\) and by current-multipole corrections \(\mathcal{M}\), helicity \(\pm1\) by a transverse current, and helicity \(0\) by the trace \(T=\eta_{\mu\nu}T^{\mu\nu}\).
Isolated sources have no radiating mass dipole; in the center-of-mass frame the conserved total momentum does not radiate.
The leading vector source is therefore a transverse current, written \(S^{j}[\dot J]\) as a Thorne placeholder, not \(\int T^{0j}\,d^{3}x\).
In GQFT that current is \(\dot{\mathbf{U}}\) of net fermion spin, as in~\eqref{eq:S1} below.
The Eardley projectors \(\Pi_{A}^{ij}\), \(\Pi_{\perp I}{}^{j}\), and \(\Pi_{\alpha}\) map these Cartesian moments onto a wave along \(\hat{\mathbf{k}}\).
Along \(\hat{\mathbf{z}}\) they reduce to
\begin{equation}
\begin{aligned}
  J_{+}
  &=
  \ddot Q_{xx}-\ddot Q_{yy}+\mathcal{M}_{+},
  &
  J_{\times}
  &=
  2\ddot Q_{xy}+\mathcal{M}_{\times},
  \\
  J_{x}
  &=
  S^{x}[\dot J],
  &
  J_{y}
  &=
  S^{y}[\dot J],
  \\
  J_{b}
  &=
  \Pi_{b}\,T,
  &
  J_{\ell}
  &=
  \Pi_{\ell}\,T,
\end{aligned}
\label{eq:Jalongz}
\end{equation}
which is Eq.~\eqref{eq:Jdef} in components.
Section~\ref{sec:eom} uses the same moments with the covariant projectors \(\Pi\).
Here \(J_{P}\) denotes the compactly supported densities whose spatial integrals reproduce~\eqref{eq:Jalongz}; these \(J_{P}\) are not the angular momentum whose derivative is \(\dot J\).
A vacuum binary has \(T=0\), so the scalar line of~\eqref{eq:eomS} remains homogeneous unless an additional scalar source is present.
Current multipoles enter \(J_{A}\) in every theory and enter \(J_{I}\) only if \(\epsV=1\).

\section{Retarded kernel}
\label{sec:green}

When \(\epsilon_{\sigma}=1\), \(\Box_{\sigma}\) is strictly hyperbolic~\cite{Maggiore:2007}.
The causal fundamental solution vanishing for \(t<0\) is
\begin{equation}
  E_{\sigma}(t,\mathbf{x})
  =
  \frac{\delta\bigl(t-|\mathbf{x}|/c_{\sigma}\bigr)}{4\pi c_{\sigma}^{2}\,|\mathbf{x}|},
  \label{eq:fund}
\end{equation}
and the unique retarded solution of \(L_{\sigma}p_{P}=\lambda_{\sigma}J_{P}\) is
\begin{equation}
  p_{P}(t,\mathbf{x})
  =
  \frac{\lambda_{\sigma}}{4\pi c_{\sigma}^{2}}
  \int
  \frac{J_{P}\bigl(t-|\mathbf{x}-\mathbf{x}'|/c_{\sigma},\,\mathbf{x}'\bigr)}{|\mathbf{x}-\mathbf{x}'|}
  \,d^{3}x'.
  \label{eq:ret}
\end{equation}
In the radiation zone \(r\gg d\), with \(r\) the distance to the source and \(d\) the source size,
\begin{equation}
\begin{gathered}
  p_{P}(t,r\hat{\mathbf{n}})
  =
  \frac{\lambda_{\sigma}}{4\pi c_{\sigma}^{2}r}\,
  \int
  J_{P}\!\left(t-\frac{r}{c_{\sigma}}+\frac{\hat{\mathbf{n}}\cdot\mathbf{x}'}{c_{\sigma}},\,\mathbf{x}'\right)
  d^{3}x'
  \\
  \qquad
  +
  O(r^{-2}),
\end{gathered}
  \label{eq:farS}
\end{equation}
an outgoing spherical wave of speed \(c_{\sigma}\).
When \(\epsilon_{\sigma}=0\) one has \(p_{P}=0\), with no retarded potential on that pair.

\section{Helmholtz and paraxial propagation}
\label{sec:helmS}

When \(\epsilon_{\sigma}=1\), \(\Box_{\sigma}\) at fixed frequency is a Helmholtz equation for the spatial amplitude of that pair.
A monochromatic wave of angular frequency \(\omega\) has spatial amplitudes \(\psi_{P}\), the Helmholtz envelopes of \(p_{P}\), distinct from a metric scalar potential, and Fourier densities \(\hat J_{P}\) of \(J_{P}\):
\begin{equation}
\begin{gathered}
  p_{P}(t,\mathbf{x})
  =
  \Re\bigl(\psi_{P}(\mathbf{x})\,e^{-i\omega t}\bigr),
  \\
  J_{P}(t,\mathbf{x})
  =
  \Re\bigl(\hat J_{P}(\mathbf{x})\,e^{-i\omega t}\bigr),
\end{gathered}
\label{eq:monoS}
\end{equation}
Substitution into Eq.~\eqref{eq:eomcomp} yields
\begin{equation}
  \bigl(\nabla^{2}+k_{P}^{2}\bigr)\,\psi_{P}
  =
  -\frac{\lambda_{\sigma}}{c_{\sigma}^{2}}\,\hat J_{P},
  \qquad
  k_{P}
  =
  \frac{\omega}{c_{\sigma}},
  \label{eq:helmS}
\end{equation}
the spatial form of \(\Box_{\sigma}\) at fixed \(\omega\), with \(\sigma\) the pair of \(P\).
Equation~\eqref{eq:helmS} is the Fourier transform of~\eqref{eq:ret}.
The tensor pair has \(c_{\mathrm{T}}=1\), so \(k_{A}=\omega\).
Vector and scalar Helmholtz equations appear only for \(\epsV=1\) and \(\epsS=1\).
The spatial amplitudes of \(\betag_{\perp}\) follow from \(\pV\) by~\eqref{eq:unifyS}.
A broadband waveform is obtained by a Fourier transform in \(t\), Helmholtz propagation at each \(\omega\) by~\eqref{eq:helmS}, and an inverse transform.

Outside the support of \(J_{P}\) the right-hand side vanishes.
Outgoing solutions have the far-field form
\begin{equation}
  \psi_{P}(\mathbf{x})
  =
  \frac{e^{ik_{P}r}}{r}\,f_{P}(\hat{\mathbf{n}}),
  \label{eq:farHS}
\end{equation}
with \(f_{P}\) an amplitude on the sky and \(\hat{\mathbf{n}}=\mathbf{x}/r\), recovering the \(1/r\) of Eq.~\eqref{eq:multipole} for the tensor pair.
A near-zone or curved-background calculation supplies \(f_{P}\) or \(\psi_{P}\) on a matching sphere;~\eqref{eq:helmS} then propagates each radiative \(p_{P}\) through the Minkowski wave zone~\cite{Isaacson:1968}.

Along a ray \(\hat{\mathbf{k}}=\hat{\mathbf{z}}\), the spatial amplitude is
\begin{equation}
  \psi_{P}(\mathbf{x})
  =
  A_{P}(\mathbf{x}_{\perp},z)\,e^{ik_{P}z},
  \label{eq:envS}
\end{equation}
with \(\mathbf{x}_{\perp}=(x,y)\) the coordinates in the transverse plane and \(A_{P}\) the envelope, distinct from the tensor index \(A\in\{+,{\times}\}\).
If \(A_{P}\) varies slowly along \(z\) compared with a wavelength, \(|\partial_{z}^{2}A_{P}|\ll k_{P}|\partial_{z}A_{P}|\),~\eqref{eq:helmS} reduces along the ray to the paraxial equation
\begin{equation}
  \nabla_{\perp}^{2}A_{P}
  +
  2ik_{P}\,\partial_{z}A_{P}
  =
  0,
  \label{eq:paraxS}
\end{equation}
where \(\nabla_{\perp}^{2}=\partial_{x}^{2}+\partial_{y}^{2}\).
Equation~\eqref{eq:paraxS} is integrated from a matching surface \(z=z_{0}\) in the radiation zone, transverse to \(\hat{\mathbf{k}}\), to a detector at \(z>z_{0}\) by the Fresnel integral:
\begin{equation}
\begin{aligned}
  A_{P}(\mathbf{x}_{\perp},z)
  &=
  \frac{k_{P}}{2\pi i(z-z_{0})}
  \int
  d^{2}x_{\perp}'\,
  A_{P}(\mathbf{x}_{\perp}',z_{0})
  \\
  &\quad\times
  \exp\Biggl(
  \frac{ik_{P}\bigl|\mathbf{x}_{\perp}-\mathbf{x}_{\perp}'\bigr|^{2}}{2(z-z_{0})}
  \Biggr).
\end{aligned}
\label{eq:fresnelS}
\end{equation}
A wave that depends only on \(u=t-z\) is the case of constant \(A_{P}\), for which \(\nabla_{\perp}^{2}A_{P}=0\).
The GEM relation~\eqref{eq:unifyS} is exact for dependence on \(u\) and holds at leading paraxial order, with corrections of relative order \(k_{P}^{-1}\bigl|\nabla_{\perp}\log A_{P}\bigr|\).
When \(c_{\sigma}\neq1\), a vector or scalar envelope diffracts on a different scale from the tensor pair.

\section{\texorpdfstring{Solution of each entry of \(\Pvec\)}{Solution of each entry of P}}
\label{sec:each}

Table~\ref{tab:eight} lists the eight couplings, the relation that solves each, and the detector that reads it.
The first six are unknowns of~\eqref{eq:eomS}; \(\betag_{\perp}\) is solved from \(\pV\).

\begin{table*}[t]
\caption{\label{tab:eight}
Eight local couplings and the model-independent relation that determines each.
Whether a given \(p_{P}\) appears in differential arm length is the separate flag \(\tau_{P}\) of~\eqref{eq:Emap}--\eqref{eq:ifo}. A Sagnac loop records GEM entries as \(\delta\ell\); an inertial sensor records \(\mathbf{a}\); a laboratory magnet records \(\boldsymbol{\Omega}\).}
\begin{ruledtabular}
\begin{tabular}{llll}
Entry & Detector coupling & Governing relation & Outgoing solution \\
\(p_{+}\) & \(\mathcal{E}_{xx}-\mathcal{E}_{yy}\), \(F_{+}\) & \(\Box p_{+}=\lambda_{\mathrm{T}}J_{+}\) & retarded~\eqref{eq:ret}; always \\
\(p_{\times}\) & \(\mathcal{E}_{xy}\), \(F_{\times}\) & \(\Box p_{\times}=\lambda_{\mathrm{T}}J_{\times}\) & retarded~\eqref{eq:ret}; always \\
\(p_{x}\) & \(\mathcal{E}_{xz}\) if \(\tau_{x}=1\) & \(L_{\mathrm{V}}p_{x}=\epsV\lambda_{\mathrm{V}}J_{x}\) & retarded if \(\epsV=1\); else \(0\) \\
\(p_{y}\) & \(\mathcal{E}_{yz}\) if \(\tau_{y}=1\) & \(L_{\mathrm{V}}p_{y}=\epsV\lambda_{\mathrm{V}}J_{y}\) & retarded if \(\epsV=1\); else \(0\) \\
\(p_{b}\) & breathing if \(\tau_{b}=1\) & \(L_{\mathrm{S}}p_{b}=\epsS\lambda_{\mathrm{S}}J_{b}\) & retarded if \(\epsS=1\); else \(0\) \\
\(p_{\ell}\) & \(\mathcal{E}_{zz}\) if \(\tau_{\ell}=1\) & \(L_{\mathrm{S}}p_{\ell}=\epsS\lambda_{\mathrm{S}}J_{\ell}\) & retarded if \(\epsS=1\); else \(0\) \\
\(\beta_{x}\) & GEM force~\eqref{eq:Fgem}; Sagnac \(\delta\ell\)\footnotemark[1] & \(\beta_{x}=-\partial_{u}p_{y}\) & image of \(p_{y}\) \\
\(\beta_{y}\) & GEM force~\eqref{eq:Fgem}; Sagnac \(\delta\ell\) & \(\beta_{y}=\partial_{u}p_{x}\) & image of \(p_{x}\) \\
\end{tabular}
\end{ruledtabular}
\footnotetext[1]{An inertial sensor or a laboratory magnet also records the GEM force.}
\end{table*}

Matching~\eqref{eq:farS} to Eq.~\eqref{eq:multipole} on the tensor pair gives
\begin{equation}
  p_{A}
  =
  \frac{2G}{c^{4}r}\,
  \Pi_{A}^{ij}
  \bigl(
  \ddot Q_{ij}
  +\mathcal{M}_{ij}(\dot J,\ddot J)
  \bigr)
  +
  O(r^{-2}),
  \label{eq:TT}
\end{equation}
with \(G\) Newton's constant, so \(p_{+}\) and \(p_{\times}\) propagate whenever \(\ddot Q_{ij}\neq0\)~\cite{LIGO:2016,Kidder:1995,Thorne:1980,Blanchet:2014}.
If \(\epsV=1\) and \(\Pi_{\perp I}{}^{j}S^{j}[\dot J]\neq0\),
\begin{equation}
  p_{I}
  =
  \frac{\lambda_{\mathrm{V}}}{4\pi c_{\mathrm{V}}^{2}r}\,
  \Pi_{\perp I}{}^{j}S^{j}[\dot J]
  +
  O(r^{-2}).
  \label{eq:Vfar}
\end{equation}
If \(\epsS=1\) and \(T\neq0\),
\begin{equation}
  p_{\alpha}
  =
  \frac{\lambda_{\mathrm{S}}}{4\pi c_{\mathrm{S}}^{2}r}\,
  \Pi_{\alpha}\,T
  +
  O(r^{-2}).
  \label{eq:Sfar}
\end{equation}
On a \(u\)-wave the GEM solution is~\eqref{eq:unifyS}; at leading order in \(1/r\),
\begin{equation}
  \betag_{\perp}(t,r\hat{\mathbf{n}})
  =
  \hat{\mathbf{n}}\times\partial_{u}\pV
  +
  O(r^{-2}).
  \label{eq:betaFar}
\end{equation}
Applying \(\partial_{u}\) and a right angle in the transverse plane to the vector line of~\eqref{eq:eomS} yields the equivalent current form \(L_{\mathrm{V}}\betag_{\perp}=\epsV\lambda_{\mathrm{V}}\mathbf{J}_{\beta}\) with \(\mathbf{J}_{\beta}=(-\partial_{u}J_{y},\partial_{u}J_{x})\). This is the same propagator acting on \(\betag_{\perp}\).

\section{Recorded time series}
\label{sec:stemp}

Equation~\eqref{eq:meas} maps \(\mathbf{p}\) and \(\betag_{\perp}\) onto the measured quantity of each coupling.
A search compares those linear responses with the data.
The predicted output of a given readout is \(s(t)\), of Fourier transform \(\tilde{s}(f)\); the arm-length case is~\eqref{eq:ifo}.
With \(S(f)\) the one-sided noise amplitude spectral density of the same readout, the detection statistic is the matched-filter signal-to-noise ratio
\begin{equation}
  \rho^{2}
  =
  4\int_{f_{\min}}^{f_{\max}}\frac{|\tilde{s}(f)|^{2}}{S(f)}\,df,
  \label{eq:rhoS}
\end{equation}
with \(f_{\min}\) and \(f_{\max}\) the edges of the observation band.
Arrival time, sky direction and the orientation of \(\pV\) are maximized over, as in a gravitational-wave template bank.
The source model that produces \(\mathbf{p}(u)\) is part of the template.
The arm-length and GEM readouts of Eq.~\eqref{eq:meas}, together with a laboratory-magnet precession template, are
\begin{equation}
  \begin{aligned}
    s_{\mathrm{IFO}}(t)
    &=
    \sum_{P:\,\tau_{P}=1}F_{P}(\hat{\mathbf{n}})\,p_{P}(t),
    \\
    s_{a}(t)
    &=
    \bigl(\mathbf{v}\times\betag_{\perp}(t)\bigr)\cdot\hat{\mathbf{e}},
    \\
    s_{\ell}(t)
    &=
    \frac{2\,\mathbf{A}\cdot\betag_{\perp}(t)}{c}\,\mathcal{F},
    \\
    s_{\Omega}(t)
    &=
    \boldsymbol{\Omega}(t)\cdot\hat{\mathbf{s}},
    \qquad
    \boldsymbol{\Omega}\propto\betag_{\perp}.
  \end{aligned}
  \label{eq:stempS}
\end{equation}
with \(\hat{\mathbf{n}}\) the source direction already used in~\eqref{eq:ifo}, \(\hat{\mathbf{e}}\) the accelerometer axis, \(\mathbf{A}\) the Sagnac vector area of the optical loop, \(\mathcal{F}\le1\) a loop form factor (unity in the long-wavelength limit), and \(\hat{\mathbf{s}}\) the axis of a laboratory-magnet readout~\cite{Mashhoon:2001}.
In the Fourier domain at a co-located detector, \(\tilde{\betag}(f)=2\pi i f\,\hat{\mathbf{k}}\times\tilde{\mathbf{p}}_{\mathrm{V}}(f)\).
The time derivative of a Michelson differential arm length \(\Delta L\), or a factor \(i\omega\) applied to a tidal spectrum, is a rewriting of the arm-length response \(s_{\mathrm{IFO}}\) and is not a GEM measurement. The GEM templates are \(s_{a}\), \(s_{\ell}\), and \(s_{\Omega}\).

A static laboratory magnetic field \(\mathbf{B}_{0}\) transduces the gravitational Larmor frequency \(\boldsymbol{\Omega}_{g}=\betag/2\).
That readout is distinct from geodesic deviation and from a Gertsenshtein conversion of a gravitational wave into a photon.
The open-circuit unipolar voltage on a disk of radius \(R_{\mathrm{B}}\) is
\begin{equation}
  s_{V}(t)
  =
  \frac14\bigl(\betag(t)\cdot\hat{\mathbf{B}}_{0}\bigr)B_{0}R_{\mathrm{B}}^{2}.
  \label{eq:Vs}
\end{equation}
Flux freezing \(\delta B/B_{0}=-\delta\mathcal{A}/\mathcal{A}\) of a circuit of area \(\mathcal{A}\) tracks electric spatial strain and vanishes when \(\tau_{x}=\tau_{y}=0\).
The voltage remains available whenever \(\epsV=1\).

\section{Plane-wave form of the eight couplings}
\label{sec:plane}

Outside the support of \(J_{P}\), Eq.~\eqref{eq:eomS} is homogeneous.
Plane waves are the homogeneous solutions of~\eqref{eq:eomS} of vanishing transverse variation.
The Fourier convention is \(e^{-i\omega t}\), as in Eq.~\eqref{eq:betaPhase}.
A common three-momentum \(\mathbf{k}\) on each sector cone has
\begin{equation}
\begin{gathered}
  \omega_{+}
  =
  \omega_{\times}
  =
  |\mathbf{k}|,
  \\
  \omega_{x}
  =
  \omega_{y}
  =
  \omega_{\mathrm{V}}
  =
  c_{\mathrm{V}}|\mathbf{k}|,
  \\
  \omega_{b}
  =
  \omega_{\ell}
  =
  \omega_{\mathrm{S}}
  =
  c_{\mathrm{S}}|\mathbf{k}|,
\end{gathered}
  \label{eq:omP}
\end{equation}
or the eigenvalues of \(M\) in place of \(\omega_{\mathrm{S}}\) if the scalars mix.
The six strain amplitudes that solve~\eqref{eq:eomS} are therefore
\begin{equation}
\begin{aligned}
  p_{+}(t,\mathbf{x})
  &=
  \Re\bigl(\hat p_{+}\,e^{-i|\mathbf{k}|t+i\mathbf{k}\cdot\mathbf{x}}\bigr),
  \\
  p_{\times}(t,\mathbf{x})
  &=
  \Re\bigl(\hat p_{\times}\,e^{-i|\mathbf{k}|t+i\mathbf{k}\cdot\mathbf{x}}\bigr),
  \\
  p_{x}(t,\mathbf{x})
  &=
  \Theta(\epsV)\,
  \Re\bigl(\hat p_{x}\,e^{-i\omega_{\mathrm{V}}t+i\mathbf{k}\cdot\mathbf{x}}\bigr),
  \\
  p_{y}(t,\mathbf{x})
  &=
  \Theta(\epsV)\,
  \Re\bigl(\hat p_{y}\,e^{-i\omega_{\mathrm{V}}t+i\mathbf{k}\cdot\mathbf{x}}\bigr),
  \\
  p_{b}(t,\mathbf{x})
  &=
  \Theta(\epsS)\,
  \Re\bigl(\hat p_{b}\,e^{-i\omega_{\mathrm{S}}t+i\mathbf{k}\cdot\mathbf{x}}\bigr),
  \\
  p_{\ell}(t,\mathbf{x})
  &=
  \Theta(\epsS)\,
  \Re\bigl(\hat p_{\ell}\,e^{-i\omega_{\mathrm{S}}t+i\mathbf{k}\cdot\mathbf{x}}\bigr).
\end{aligned}
\label{eq:pPlane}
\end{equation}
Each line satisfies \(\bigl(-\omega_{P}^{2}+c_{P}^{2}|\mathbf{k}|^{2}\bigr)\hat p_{P}=0\) on its cone, and vanishes identically if that pair is constrained.
The GEM pair follows from~\eqref{eq:unifyS} applied to~\eqref{eq:pPlane}.
Along the vector ray, \(u=t-\hat{\mathbf{k}}\cdot\mathbf{x}/c_{\mathrm{V}}\) and \(\partial_{u}\) multiplies Fourier amplitudes by \(-i\omega_{\mathrm{V}}\), so
\begin{equation}
\begin{aligned}
  \beta_{x}(t,\mathbf{x})
  &=
  \Theta(\epsV)\,
  \Re\bigl(i\omega_{\mathrm{V}}\,\hat p_{y}\,e^{-i\omega_{\mathrm{V}}t+i\mathbf{k}\cdot\mathbf{x}}\bigr),
  \\
  \beta_{y}(t,\mathbf{x})
  &=
  \Theta(\epsV)\,
  \Re\bigl(-i\omega_{\mathrm{V}}\,\hat p_{x}\,e^{-i\omega_{\mathrm{V}}t+i\mathbf{k}\cdot\mathbf{x}}\bigr),
\end{aligned}
\label{eq:betaPlane}
\end{equation}
which is Eq.~\eqref{eq:betaPhase}.
The eight couplings on a single \(\mathbf{k}\) are therefore the real part of
\begin{equation}
\begin{gathered}
  \hat{\Pvec}(\mathbf{k})
  =
  \bigl(
  \hat p_{+},\,
  \hat p_{\times},\,
  \Theta(\epsV)\hat p_{x},\,
  \Theta(\epsV)\hat p_{y},
  \\
  \qquad
  \Theta(\epsS)\hat p_{b},\,
  \Theta(\epsS)\hat p_{\ell},
  \\
  \qquad
  \Theta(\epsV)\,i\omega_{\mathrm{V}}\hat p_{y},\,
  \Theta(\epsV)\,(-i\omega_{\mathrm{V}})\hat p_{x}
  \bigr).
\end{gathered}
\label{eq:Phat}
\end{equation}
with the tensor pair oscillating at \(|\mathbf{k}|\), the vector and GEM entries at \(\omega_{\mathrm{V}}\), and the scalar pair at \(\omega_{\mathrm{S}}\).
GEM occupies the same exponential as \(\pV\), never the tensor or scalar cone, in agreement with~\eqref{eq:unifyS}.
If \(\epsV=0\), the last two entries of~\eqref{eq:Phat} vanish with \(\hat p_{x}\) and \(\hat p_{y}\).

On a wave along \(\hat{\mathbf{z}}\), Eq.~\eqref{eq:pPlane} depends only on the sector retarded time, and~\eqref{eq:Emap} becomes \(\mathcal{E}_{ij}\propto-\omega_{P}^{2}p_{P}\) for each tidal entry with \(\tau_{P}=1\).
For linear vector polarization of amplitude \(a\) and phase \(\phi\), \(p_{x}=a\cos(\omega_{\mathrm{V}}u+\phi)\), \(p_{y}=0\), Eq.~\eqref{eq:betaPlane} gives \(\beta_{x}=0\) and \(\beta_{y}=-\omega_{\mathrm{V}}a\sin(\omega_{\mathrm{V}}u+\phi)\), and the intensities alternate,
\begin{equation}
  |\pV|^{2}
  +
  \omega_{\mathrm{V}}^{-2}|\betag_{\perp}|^{2}
  =
  a^{2}.
  \label{eq:altS}
\end{equation}
A general homogeneous solution is a superposition of~\eqref{eq:pPlane}--\eqref{eq:betaPlane} over \(\mathbf{k}\) on each active cone.

\section{\texorpdfstring{GQFT row: \(\dot{\mathbf{U}}\neq0\) and GEM measurements}{GQFT row: vector source and GEM measurements}}
\label{sec:gqft}

The GQFT row of Table~\ref{tab:models} is the existence example, used in Sec.~\ref{sec:faraday}, that a GEM wave can sit in the mixture without a geodesic-deviation vector, equivalently without an arm-length vector channel.
It is read from the linearized radiation of Gao \emph{et al.}~\cite{Gao:2025GW}.
In that realization the metric vector potentials are \(h_{0i}=S_{i}\) and \(h_{ij}=2\partial_{(i}F_{j)}\), with \(\partial^{i}S_{i}=\partial^{i}F_{i}=0\).
These \(S_{i}\) and \(F_{i}\) are distinct from the Thorne current \(S^{j}[\dot J]\) of~\eqref{eq:Jalongz}, the GEM force \(F^{i}\) of~\eqref{eq:Fgem}, and the antenna patterns \(F_{P}\).
The wave operator \(L_{\sigma}\) of~\eqref{eq:LsigS} is distinct from the orbital angular momentum \(\mathbf{L}_{\mathrm{orb}}\) below, and the source densities \(J_{P}\) are distinct from \(\dot J\) and from the net fermion spin \(\mathbf{U}\).
Retarded time on a ray is \(u\).

Orbital angular momentum \(\mathbf{L}_{\mathrm{orb}}=\int\varepsilon^{ijk}y_{j}(v_{k}+u_{k})\,d^{3}y\) is conserved, \(d\mathbf{L}_{\mathrm{orb}}/dt=0\), with \(v_{k}\) and \(u_{k}\) the transverse scalar-vector-tensor (SVT) projections of the symmetric and antisymmetric stresses \(T_{(0k)}\) and \(T_{[0k]}\), distinct from retarded time \(u\)~\cite{Gao:2025GW}.
It supplies a near-zone field \(S_{i}^{(L)}\propto\varepsilon_{ijk}\hat{k}^{j}L_{\mathrm{orb}}^{k}/r^{2}\), which is the Lense--Thirring piece excluded from radiative \(\Pvec\) as \(\omega\to0\).
The fermion spin density is \(\Xi^{k}\).
The net spin is \(U^{k}=-\int\Xi^{k}\,d^{3}y\), and \(t_{r}=t-r\) is retarded time at distance \(r\).
The radiating \(1/r\) vector field is
\begin{equation}
  S_{i}^{(1)}
  \propto
  \varepsilon_{ijk}\hat{k}^{j}\,\frac{dU^{k}}{dt_{r}}\,,
  \label{eq:S1}
\end{equation}
so a nonzero helicity-\(\pm1\) current in GQFT is \(\dot{\mathbf{U}}\neq0\), rather than orbital \(\dot{\mathbf{L}}_{\mathrm{orb}}\).
The placeholder \(S^{j}[\dot J]\) of~\eqref{eq:Jalongz} is realized by this spin current.

On a source-free \(u\)-wave, \(S_{i}=\partial_{t}F_{i}\)~\cite{Gao:2025GW}.
The two radiative functions are the transverse pair \(S_{\perp}\) or equivalently \(F_{\perp}\).
The \(\pV\) of Eq.~\eqref{eq:pdef} is identified with this transverse pair, \(h_{0i}=S_{i}\), and~\eqref{eq:unifyS} gives
\begin{equation}
  \betag_{\perp}
  =
  \hat{\mathbf{k}}\times\partial_{u}\mathbf{S}_{\perp},
  \qquad
  \beta_{x}=-\partial_{u}S_{y},
  \qquad
  \beta_{y}=\partial_{u}S_{x}
  \label{eq:betaS}
\end{equation}
along \(\hat{\mathbf{z}}\).
Whenever \(S_{i}^{(1)}\) oscillates, both \(\beta_{x}\) and \(\beta_{y}\) are nonzero radiative measurements: they are the GEM components in~\eqref{eq:Fgem}, recorded by a Sagnac loop.
The laboratory magnet of~\eqref{eq:Vs} transduces that same \(\boldsymbol{\Omega}_{g}=\betag/2\).
Flux freezing \(\delta B/B_{0}=-\delta\mathcal{A}/\mathcal{A}\) of a circuit of area \(\mathcal{A}\) tracks electric spatial strain and is off when \(\tau_{x}=\tau_{y}=0\), as is the arm-length \(\Delta L\) of the vector sector.
The GEM readout is then a voltage at \(\omega|\mathbf{S}_{\perp}|\) with no accompanying \(\delta B\) and no Eardley vector strain.

The kinematic map~\eqref{eq:unifyS} still holds: \(\betag_{\perp}\) is one retarded derivative of \(\pV\).
On a monochromatic wave that derivative is a phase \(\pi/2\), Eq.~\eqref{eq:betaPhase}, and the intensities alternate,~\eqref{eq:altS}; on a broadband waveform it is the local rate of change of \(\pV\).
That relation does not make \(\betag_{\perp}\) an arm-length observable.
Rest-mass Michelson combinations operate with essentially rest masses, \(v\approx0\), and read geodesic deviation~\eqref{eq:geoS}, hence \(\mathcal{E}_{ij}\sim R_{i00j}\), not the GEM force~\eqref{eq:Fgem}.
Even when \(\tau_{x}=\tau_{y}=1\), geodesic deviation would record \(\partial_{u}^{2}\pV=\partial_{u}\betag_{\perp}\), a further derivative, not \(\betag_{\perp}\) itself.
Geodesic deviation is controlled by \(R_{i00j}=-\partial_{t}\partial_{(i}[S-\partial_{t}F]_{j)}\).
On the source-free cone this vanishes, so \(\tau_{x}=\tau_{y}=0\): neither \(\pV\) nor \(\partial_{u}^{2}\pV\) appears in~\eqref{eq:ifo}~\cite{Gao:2025GW}.
The \(\pi/2\) partner of \(\pV\) is therefore a GEM measurement of the rate of change of \(S_{\perp}\), not a phase-shifted copy of an arm-length template.
The same two radiative functions still enter \(\Nrad\) and still determine \(\betag_{\perp}\).

An explicit source with \(\dot{\mathbf{U}}\neq0\) is an equal-mass neutron-star binary in which one star's net spin tracks its orbital velocity, \(-U^{i}_{(1)}=U_{0}\hat{v}^{i}(t)\), with \(U_{0}\) the constant amplitude of that spin~\cite{Gao:2025GW}.
The centre-of-mass position of the spinning star is \(\mathbf{R}_{(1)}\), of orbital radius \(R\).
Then \(S_{i}^{(1)}\propto(\omega U_{0}/(Rr))\,\varepsilon_{ijk}\hat{k}^{j}R_{(1)}^{k}(t_{r})\) oscillates at the orbital angular frequency \(\omega\).
Along \(\hat{\mathbf{z}}\),~\eqref{eq:betaS} yields \(\beta_{x}\) and \(\beta_{y}\) as explicit, nonzero functions of retarded time, while \(\mathcal{E}_{xz}=\mathcal{E}_{yz}=0\).
This is the GQFT row of Table~\ref{tab:Nrad}: a time-varying helicity-\(\pm1\) current with \(\epsV=1\) feeds \(\pV\) and \(\betag_{\perp}\), the arm-length vector channel is off, and the measurable pair is \((\beta_{x},\beta_{y})\), recorded by a Sagnac loop.

\section{Dictionary for Table~I and for other models}
\label{sec:dict}

A linearized theory in the Minkowski radiation zone is completely specified, for Table~\ref{tab:Nrad}, by the discrete data
\begin{equation}
  \bigl(\epsV,\,\epsS,\,n_{\mathrm{S}},\,\tau_{P},\,c_{\sigma}\bigr)
  \label{eq:data}
\end{equation}
together with which moments of \(T^{\mu\nu}\) are available as sources.
The switches \((\epsV,\epsS)\) select propagating versus constrained pairs in~\eqref{eq:eomS}.
The integer \(n_{\mathrm{S}}\) is the number of independent scalar eigenmodes of \(M\).
The flags \(\tau_{P}\) select which propagating amplitudes enter \(\mathcal{E}_{ij}\) and therefore~\eqref{eq:ifo}.
The speeds \(c_{\sigma}\) label the cones; only \(c_{\mathrm{T}}=1\) is fixed by GW170817~\cite{LIGO:2017GW170817}.
Table~\ref{tab:Nrad} lists \((\epsV,\epsS)\), \(\Nrad\), the arm-length entries with \(\tau_{P}=1\), and whether GEM is present; Table~\ref{tab:models} expands that list by \(n_{\mathrm{S}}\) and notes, so that another model is read by filling the same columns~\cite{Brans:1961,Jacobson:2004,Jacobson:2007,Will:2014,Dong:2024,Gao:2025GW}.
Once those columns are filled, the measured quantity of each coupling isolates the polarizations in the mixture.

\begin{table*}[t]
\caption{\label{tab:models}
Detail of Table~\ref{tab:Nrad}.
\(\Nrad\) counts outgoing strain branches, Eq.~\eqref{eq:NradS}.
The interferometer (IFO) column lists the amplitudes that enter differential arm length (\(\tau_{P}=1\)).
GEM is present whenever \(\epsV=1\), independently of \(\tau_{x},\tau_{y}\).
Each row records which time series that assignment predicts.}
\begin{ruledtabular}
\begin{tabular}{lcccccl}
Theory & \((\epsV,\epsS)\) & \(n_{\mathrm{S}}\) & \(\Nrad\) & IFO (\(\tau_{P}=1\)) & GEM & Notes \\
GR & \((0,0)\) & \(0\) & \(2\) & \(p_{+},p_{\times}\) & \(0\) & Einstein constraints \\
scalar-tensor & \((0,1)\) & \(1\) & \(3\) & \(\pT\), one scalar & \(0\) & typically \(p_{b}\)~\cite{Brans:1961,Will:2014} \\
Einstein--{\AE}ther & \((1,1)\) & \(1\) & \(5\) & \(\pT\), \(\pV\), one scalar & \(\betag_{\perp}\) & \(\tau_{x}=\tau_{y}=1\)~\cite{Jacobson:2004} \\
GQFT & \((1,1)\) & \(1\) & \(5\) & \(\pT\), \(p_{b}\) & \(\betag_{\perp}\) & \(\tau_{x}=\tau_{y}=0\)~\cite{Gao:2025GW} \\
maximal & \((1,1)\) & \(1\) or \(2\) & \(\le6\) & all \(p_{P}\) with \(\tau_{P}=1\) & \(\betag_{\perp}\) & Dong~\cite{Dong:2024} \\
\end{tabular}
\end{ruledtabular}
\end{table*}

The rows are obtained as follows.
GR imposes \(\epsV=\epsS=0\), so \(p_{I}=p_{\alpha}=0\) and~\eqref{eq:unifyS} gives \(\betag_{\perp}=0\); only~\eqref{eq:TT} remains~\cite{Maggiore:2007}.
Scalar-tensor theories keep \(\epsV=0\) and turn on one scalar eigenmode, usually breathing, sourced by \(T\)~\cite{Brans:1961,Will:2014}.
Einstein--{\AE}ther has five propagating modes whose vector pair does enter \(\mathcal{E}_{ij}\), so differential arm length records \(\pV\) and a Sagnac loop records \(\partial_{u}\pV\)~\cite{Jacobson:2004,Jacobson:2007}.
Rest-mass Michelson combinations never record the GEM force itself: that is true for every row with \(\epsV=1\).
GQFT has five propagating modes as well, but the two vector modes are transverse in the scalar-vector-tensor decomposition and do not enter \(\mathcal{E}_{ij}\): \(\Nrad\) includes them, Eq.~\eqref{eq:ifo} does not, and \(\betag_{\perp}\) need not vanish~\cite{Gao:2025GW}.
Equations~\eqref{eq:S1}--\eqref{eq:betaS} give \(\beta_{x}\) and \(\beta_{y}\) as the vector measurements when \(\dot{\mathbf{U}}\neq0\).
The maximal assignment is \(\epsV=\epsS=1\) with \(n_{\mathrm{S}}\le2\), which is the Eardley upper bound of six strain polarizations.
Dong's classification of metric and scalar-tensor theories reaches that bound in some parameter regions, while forbidding vector modes when the tensors propagate only at the speed of light~\cite{Dong:2024}; the measured subset is whichever \(\tau_{P}\) a given realization sets to one.

Another model is placed on Table~\ref{tab:models} by linearizing in the Minkowski radiation zone; the operator~\eqref{eq:eomS} need not be rederived.
(i) The propagating field decomposes in the \(E(2)\) basis \(\mathbf{p}\).
(ii) The principal symbol on each pair is a null cone, which gives \(\epsilon_{\sigma}=1\) and a speed \(c_{\sigma}\), or a constraint, which gives \(\epsilon_{\sigma}=0\).
(iii) The tidal tensor \(\mathcal{E}_{ij}\) sets \(\tau_{P}=1\) if and only if that \(p_{P}\) appears in~\eqref{eq:Emap}.
(iv) The scalar kinetic matrix yields \(n_{\mathrm{S}}\), after which~\eqref{eq:NradS} follows.
The retarded solutions~\eqref{eq:TT}--\eqref{eq:betaFar} and the plane-wave solutions~\eqref{eq:pPlane}--\eqref{eq:betaPlane} then hold with the sources of that model inserted into~\eqref{eq:Jalongz}.
Two models with the same \((\epsV,\epsS,n_{\mathrm{S}})\) may still differ in \(\tau_{P}\), as Einstein--{\AE}ther and GQFT illustrate: they share \(\Nrad=5\) and differ in whether the vector sector appears in arm length or only in a GEM readout.

\bibliographystyle{apsrev4-1}
\bibliography{polar}

\begin{thebibliography}{25}%
\makeatletter
\providecommand \@ifxundefined [1]{%
 \@ifx{#1\undefined}
}%
\providecommand \@ifnum [1]{%
 \ifnum #1\expandafter \@firstoftwo
 \else \expandafter \@secondoftwo
 \fi
}%
\providecommand \@ifx [1]{%
 \ifx #1\expandafter \@firstoftwo
 \else \expandafter \@secondoftwo
 \fi
}%
\providecommand \natexlab [1]{#1}%
\providecommand \enquote  [1]{``#1''}%
\providecommand \bibnamefont  [1]{#1}%
\providecommand \bibfnamefont [1]{#1}%
\providecommand \citenamefont [1]{#1}%
\providecommand \href@noop [0]{\@secondoftwo}%
\providecommand \href [0]{\begingroup \@sanitize@url \@href}%
\providecommand \@href[1]{\@@startlink{#1}\@@href}%
\providecommand \@@href[1]{\endgroup#1\@@endlink}%
\providecommand \@sanitize@url [0]{\catcode `\\12\catcode `\$12\catcode
  `\&12\catcode `\#12\catcode `\^12\catcode `\_12\catcode `\%12\relax}%
\providecommand \@@startlink[1]{}%
\providecommand \@@endlink[0]{}%
\providecommand \url  [0]{\begingroup\@sanitize@url \@url }%
\providecommand \@url [1]{\endgroup\@href {#1}{\urlprefix }}%
\providecommand \urlprefix  [0]{URL }%
\providecommand \Eprint [0]{\href }%
\providecommand \doibase [0]{http://dx.doi.org/}%
\providecommand \selectlanguage [0]{\@gobble}%
\providecommand \bibinfo  [0]{\@secondoftwo}%
\providecommand \bibfield  [0]{\@secondoftwo}%
\providecommand \translation [1]{[#1]}%
\providecommand \BibitemOpen [0]{}%
\providecommand \bibitemStop [0]{}%
\providecommand \bibitemNoStop [0]{.\EOS\space}%
\providecommand \EOS [0]{\spacefactor3000\relax}%
\providecommand \BibitemShut  [1]{\csname bibitem#1\endcsname}%
\let\auto@bib@innerbib\@empty
\bibitem [{\citenamefont {Abbott}\ \emph {et~al.}(2016)\citenamefont {Abbott}
  \emph {et~al.}}]{LIGO:2016}%
  \BibitemOpen
  \bibfield  {author} {\bibinfo {author} {\bibfnamefont {B.~P.}\ \bibnamefont
  {Abbott}} \emph {et~al.} (\bibinfo {collaboration} {LIGO Scientific
  Collaboration and Virgo Collaboration}),\ }\href {\doibase
  10.1103/PhysRevLett.116.061102} {\bibfield  {journal} {\bibinfo  {journal}
  {Phys. Rev. Lett.}\ }\textbf {\bibinfo {volume} {116}},\ \bibinfo {pages}
  {061102} (\bibinfo {year} {2016})},\ \Eprint
  {http://arxiv.org/abs/1602.03837} {arXiv:1602.03837} \BibitemShut {NoStop}%
\bibitem [{\citenamefont {Nishizawa}\ \emph {et~al.}(2009)\citenamefont
  {Nishizawa}, \citenamefont {Taruya}, \citenamefont {Hayama}, \citenamefont
  {Kawamura},\ and\ \citenamefont {Sakagami}}]{Nishizawa:2009}%
  \BibitemOpen
  \bibfield  {author} {\bibinfo {author} {\bibfnamefont {A.}~\bibnamefont
  {Nishizawa}}, \bibinfo {author} {\bibfnamefont {A.}~\bibnamefont {Taruya}},
  \bibinfo {author} {\bibfnamefont {K.}~\bibnamefont {Hayama}}, \bibinfo
  {author} {\bibfnamefont {S.}~\bibnamefont {Kawamura}}, \ and\ \bibinfo
  {author} {\bibfnamefont {M.-a.}\ \bibnamefont {Sakagami}},\ }\href {\doibase
  10.1103/PhysRevD.79.082002} {\bibfield  {journal} {\bibinfo  {journal} {Phys.
  Rev. D}\ }\textbf {\bibinfo {volume} {79}},\ \bibinfo {pages} {082002}
  (\bibinfo {year} {2009})},\ \Eprint {http://arxiv.org/abs/0903.0528}
  {arXiv:0903.0528} \BibitemShut {NoStop}%
\bibitem [{\citenamefont {Hayama}\ and\ \citenamefont
  {Nishizawa}(2013)}]{Hayama:2013}%
  \BibitemOpen
  \bibfield  {author} {\bibinfo {author} {\bibfnamefont {K.}~\bibnamefont
  {Hayama}}\ and\ \bibinfo {author} {\bibfnamefont {A.}~\bibnamefont
  {Nishizawa}},\ }\href {\doibase 10.1103/PhysRevD.87.062003} {\bibfield
  {journal} {\bibinfo  {journal} {Phys. Rev. D}\ }\textbf {\bibinfo {volume}
  {87}},\ \bibinfo {pages} {062003} (\bibinfo {year} {2013})},\ \Eprint
  {http://arxiv.org/abs/1208.4596} {arXiv:1208.4596} \BibitemShut {NoStop}%
\bibitem [{\citenamefont {Isi}\ \emph {et~al.}(2017)\citenamefont {Isi},
  \citenamefont {Pitkin},\ and\ \citenamefont {Weinstein}}]{Isi:2017}%
  \BibitemOpen
  \bibfield  {author} {\bibinfo {author} {\bibfnamefont {M.}~\bibnamefont
  {Isi}}, \bibinfo {author} {\bibfnamefont {M.}~\bibnamefont {Pitkin}}, \ and\
  \bibinfo {author} {\bibfnamefont {A.~J.}\ \bibnamefont {Weinstein}},\ }\href
  {\doibase 10.1103/PhysRevD.96.042001} {\bibfield  {journal} {\bibinfo
  {journal} {Phys. Rev. D}\ }\textbf {\bibinfo {volume} {96}},\ \bibinfo
  {pages} {042001} (\bibinfo {year} {2017})},\ \Eprint
  {http://arxiv.org/abs/1703.07530} {arXiv:1703.07530} \BibitemShut {NoStop}%
\bibitem [{\citenamefont {Takeda}\ \emph {et~al.}(2018)\citenamefont {Takeda},
  \citenamefont {Nishizawa}, \citenamefont {Michimura}, \citenamefont {Nagano},
  \citenamefont {Komori}, \citenamefont {Ando},\ and\ \citenamefont
  {Hayama}}]{Takeda:2018}%
  \BibitemOpen
  \bibfield  {author} {\bibinfo {author} {\bibfnamefont {H.}~\bibnamefont
  {Takeda}}, \bibinfo {author} {\bibfnamefont {A.}~\bibnamefont {Nishizawa}},
  \bibinfo {author} {\bibfnamefont {Y.}~\bibnamefont {Michimura}}, \bibinfo
  {author} {\bibfnamefont {K.}~\bibnamefont {Nagano}}, \bibinfo {author}
  {\bibfnamefont {K.}~\bibnamefont {Komori}}, \bibinfo {author} {\bibfnamefont
  {M.}~\bibnamefont {Ando}}, \ and\ \bibinfo {author} {\bibfnamefont
  {K.}~\bibnamefont {Hayama}},\ }\href {\doibase 10.1103/PhysRevD.98.022008}
  {\bibfield  {journal} {\bibinfo  {journal} {Phys. Rev. D}\ }\textbf {\bibinfo
  {volume} {98}},\ \bibinfo {pages} {022008} (\bibinfo {year} {2018})},\
  \Eprint {http://arxiv.org/abs/1806.02182} {arXiv:1806.02182} \BibitemShut
  {NoStop}%
\bibitem [{\citenamefont {Maggiore}(2007)}]{Maggiore:2007}%
  \BibitemOpen
  \bibfield  {author} {\bibinfo {author} {\bibfnamefont {M.}~\bibnamefont
  {Maggiore}},\ }\href@noop {} {\emph {\bibinfo {title} {Gravitational Waves.
  {V}ol. 1: {T}heory and Experiments}}}\ (\bibinfo  {publisher} {Oxford
  University Press},\ \bibinfo {year} {2007})\BibitemShut {NoStop}%
\bibitem [{\citenamefont {Eardley}\ \emph
  {et~al.}(1973{\natexlab{a}})\citenamefont {Eardley}, \citenamefont {Lee},
  \citenamefont {Lightman}, \citenamefont {Wagoner},\ and\ \citenamefont
  {Will}}]{Eardley:1973}%
  \BibitemOpen
  \bibfield  {author} {\bibinfo {author} {\bibfnamefont {D.~M.}\ \bibnamefont
  {Eardley}}, \bibinfo {author} {\bibfnamefont {D.~L.}\ \bibnamefont {Lee}},
  \bibinfo {author} {\bibfnamefont {A.~P.}\ \bibnamefont {Lightman}}, \bibinfo
  {author} {\bibfnamefont {R.~V.}\ \bibnamefont {Wagoner}}, \ and\ \bibinfo
  {author} {\bibfnamefont {C.~M.}\ \bibnamefont {Will}},\ }\href {\doibase
  10.1103/PhysRevLett.30.884} {\bibfield  {journal} {\bibinfo  {journal} {Phys.
  Rev. Lett.}\ }\textbf {\bibinfo {volume} {30}},\ \bibinfo {pages} {884}
  (\bibinfo {year} {1973}{\natexlab{a}})}\BibitemShut {NoStop}%
\bibitem [{\citenamefont {Eardley}\ \emph
  {et~al.}(1973{\natexlab{b}})\citenamefont {Eardley}, \citenamefont {Lee},\
  and\ \citenamefont {Lightman}}]{Eardley:1973PRD}%
  \BibitemOpen
  \bibfield  {author} {\bibinfo {author} {\bibfnamefont {D.~M.}\ \bibnamefont
  {Eardley}}, \bibinfo {author} {\bibfnamefont {D.~L.}\ \bibnamefont {Lee}}, \
  and\ \bibinfo {author} {\bibfnamefont {A.~P.}\ \bibnamefont {Lightman}},\
  }\href {\doibase 10.1103/PhysRevD.8.3308} {\bibfield  {journal} {\bibinfo
  {journal} {Phys. Rev. D}\ }\textbf {\bibinfo {volume} {8}},\ \bibinfo {pages}
  {3308} (\bibinfo {year} {1973}{\natexlab{b}})}\BibitemShut {NoStop}%
\bibitem [{\citenamefont {Ruggiero}\ and\ \citenamefont
  {Ortolan}(2020{\natexlab{a}})}]{Ruggiero:2020GEM}%
  \BibitemOpen
  \bibfield  {author} {\bibinfo {author} {\bibfnamefont {M.~L.}\ \bibnamefont
  {Ruggiero}}\ and\ \bibinfo {author} {\bibfnamefont {A.}~\bibnamefont
  {Ortolan}},\ }\href {\doibase 10.1088/2399-6528/ab9320} {\bibfield  {journal}
  {\bibinfo  {journal} {J. Phys. Commun.}\ }\textbf {\bibinfo {volume} {4}},\
  \bibinfo {pages} {055013} (\bibinfo {year} {2020}{\natexlab{a}})},\ \Eprint
  {http://arxiv.org/abs/2003.05207} {arXiv:2003.05207} \BibitemShut {NoStop}%
\bibitem [{\citenamefont {Ruggiero}\ and\ \citenamefont
  {Ortolan}(2020{\natexlab{b}})}]{Ruggiero:2020res}%
  \BibitemOpen
  \bibfield  {author} {\bibinfo {author} {\bibfnamefont {M.~L.}\ \bibnamefont
  {Ruggiero}}\ and\ \bibinfo {author} {\bibfnamefont {A.}~\bibnamefont
  {Ortolan}},\ }\href {\doibase 10.1103/PhysRevD.102.101501} {\bibfield
  {journal} {\bibinfo  {journal} {Phys. Rev. D}\ }\textbf {\bibinfo {volume}
  {102}},\ \bibinfo {pages} {101501} (\bibinfo {year} {2020}{\natexlab{b}})},\
  \Eprint {http://arxiv.org/abs/2011.01663} {arXiv:2011.01663} \BibitemShut
  {NoStop}%
\bibitem [{\citenamefont {Mashhoon}(2001)}]{Mashhoon:2001}%
  \BibitemOpen
  \bibfield  {author} {\bibinfo {author} {\bibfnamefont {B.}~\bibnamefont
  {Mashhoon}},\ }in\ \href {\doibase 10.1142/9789812810021_0009} {\emph
  {\bibinfo {booktitle} {Reference Frames and Gravitomagnetism}}},\ \bibinfo
  {editor} {edited by\ \bibinfo {editor} {\bibfnamefont {J.-F.}\ \bibnamefont
  {Pascual-S{\'a}nchez}}, \bibinfo {editor} {\bibfnamefont {L.}~\bibnamefont
  {Flor{\'\i}a}}, \bibinfo {editor} {\bibfnamefont {A.}~\bibnamefont
  {San~Miguel}}, \ and\ \bibinfo {editor} {\bibfnamefont {F.}~\bibnamefont
  {Vicente}}}\ (\bibinfo  {publisher} {World Scientific},\ \bibinfo {year}
  {2001})\ \Eprint {http://arxiv.org/abs/gr-qc/0011014} {arXiv:gr-qc/0011014}
  \BibitemShut {NoStop}%
\bibitem [{\citenamefont {Brans}\ and\ \citenamefont
  {Dicke}(1961)}]{Brans:1961}%
  \BibitemOpen
  \bibfield  {author} {\bibinfo {author} {\bibfnamefont {C.}~\bibnamefont
  {Brans}}\ and\ \bibinfo {author} {\bibfnamefont {R.~H.}\ \bibnamefont
  {Dicke}},\ }\href {\doibase 10.1103/PhysRev.124.925} {\bibfield  {journal}
  {\bibinfo  {journal} {Phys. Rev.}\ }\textbf {\bibinfo {volume} {124}},\
  \bibinfo {pages} {925} (\bibinfo {year} {1961})}\BibitemShut {NoStop}%
\bibitem [{\citenamefont {Jacobson}\ and\ \citenamefont
  {Mattingly}(2004)}]{Jacobson:2004}%
  \BibitemOpen
  \bibfield  {author} {\bibinfo {author} {\bibfnamefont {T.}~\bibnamefont
  {Jacobson}}\ and\ \bibinfo {author} {\bibfnamefont {D.}~\bibnamefont
  {Mattingly}},\ }\href {\doibase 10.1103/PhysRevD.70.024003} {\bibfield
  {journal} {\bibinfo  {journal} {Phys. Rev. D}\ }\textbf {\bibinfo {volume}
  {70}},\ \bibinfo {pages} {024003} (\bibinfo {year} {2004})},\ \Eprint
  {http://arxiv.org/abs/gr-qc/0402005} {arXiv:gr-qc/0402005} \BibitemShut
  {NoStop}%
\bibitem [{\citenamefont {Will}(2014)}]{Will:2014}%
  \BibitemOpen
  \bibfield  {author} {\bibinfo {author} {\bibfnamefont {C.~M.}\ \bibnamefont
  {Will}},\ }\href {\doibase 10.12942/lrr-2014-4} {\bibfield  {journal}
  {\bibinfo  {journal} {Living Rev. Rel.}\ }\textbf {\bibinfo {volume} {17}},\
  \bibinfo {pages} {4} (\bibinfo {year} {2014})},\ \Eprint
  {http://arxiv.org/abs/1403.7377} {arXiv:1403.7377} \BibitemShut {NoStop}%
\bibitem [{\citenamefont {Dong}\ \emph {et~al.}(2024)\citenamefont {Dong},
  \citenamefont {Liu},\ and\ \citenamefont {Liu}}]{Dong:2024}%
  \BibitemOpen
  \bibfield  {author} {\bibinfo {author} {\bibfnamefont {Y.-Q.}\ \bibnamefont
  {Dong}}, \bibinfo {author} {\bibfnamefont {Y.-Q.}\ \bibnamefont {Liu}}, \
  and\ \bibinfo {author} {\bibfnamefont {Y.-X.}\ \bibnamefont {Liu}},\ }\href
  {\doibase 10.1103/PhysRevD.109.044013} {\bibfield  {journal} {\bibinfo
  {journal} {Phys. Rev. D}\ }\textbf {\bibinfo {volume} {109}},\ \bibinfo
  {pages} {044013} (\bibinfo {year} {2024})},\ \Eprint
  {http://arxiv.org/abs/2310.11336} {arXiv:2310.11336} \BibitemShut {NoStop}%
\bibitem [{\citenamefont {Gao}\ \emph {et~al.}(2025)\citenamefont {Gao},
  \citenamefont {Huang},\ and\ \citenamefont {Wu}}]{Gao:2025GW}%
  \BibitemOpen
  \bibfield  {author} {\bibinfo {author} {\bibfnamefont {Y.-K.}\ \bibnamefont
  {Gao}}, \bibinfo {author} {\bibfnamefont {D.}~\bibnamefont {Huang}}, \ and\
  \bibinfo {author} {\bibfnamefont {Y.-L.}\ \bibnamefont {Wu}},\ }\href
  {\doibase 10.1140/epjc/s10052-025-14889-1} {\bibfield  {journal} {\bibinfo
  {journal} {Eur. Phys. J. C}\ }\textbf {\bibinfo {volume} {85}},\ \bibinfo
  {pages} {1159} (\bibinfo {year} {2025})},\ \Eprint
  {http://arxiv.org/abs/2506.21225} {arXiv:2506.21225} \BibitemShut {NoStop}%
\bibitem [{\citenamefont {Isaacson}(1968)}]{Isaacson:1968}%
  \BibitemOpen
  \bibfield  {author} {\bibinfo {author} {\bibfnamefont {R.~A.}\ \bibnamefont
  {Isaacson}},\ }\href {\doibase 10.1103/PhysRev.166.1263} {\bibfield
  {journal} {\bibinfo  {journal} {Phys. Rev.}\ }\textbf {\bibinfo {volume}
  {166}},\ \bibinfo {pages} {1263} (\bibinfo {year} {1968})}\BibitemShut
  {NoStop}%
\bibitem [{\citenamefont {Thorne}(1980)}]{Thorne:1980}%
  \BibitemOpen
  \bibfield  {author} {\bibinfo {author} {\bibfnamefont {K.~S.}\ \bibnamefont
  {Thorne}},\ }\href {\doibase 10.1103/RevModPhys.52.299} {\bibfield  {journal}
  {\bibinfo  {journal} {Rev. Mod. Phys.}\ }\textbf {\bibinfo {volume} {52}},\
  \bibinfo {pages} {299} (\bibinfo {year} {1980})}\BibitemShut {NoStop}%
\bibitem [{\citenamefont {Abbott}\ \emph
  {et~al.}(2017{\natexlab{a}})\citenamefont {Abbott} \emph
  {et~al.}}]{LIGO:2017GW170817}%
  \BibitemOpen
  \bibfield  {author} {\bibinfo {author} {\bibfnamefont {B.~P.}\ \bibnamefont
  {Abbott}} \emph {et~al.} (\bibinfo {collaboration} {LIGO Scientific
  Collaboration and Virgo Collaboration and Fermi-GBM and INTEGRAL}),\ }\href
  {\doibase 10.3847/2041-8213/aa920c} {\bibfield  {journal} {\bibinfo
  {journal} {Astrophys. J. Lett.}\ }\textbf {\bibinfo {volume} {848}},\
  \bibinfo {pages} {L13} (\bibinfo {year} {2017}{\natexlab{a}})},\ \Eprint
  {http://arxiv.org/abs/1710.05834} {arXiv:1710.05834} \BibitemShut {NoStop}%
\bibitem [{\citenamefont {Ciufolini}(2000)}]{Ciufolini:2000}%
  \BibitemOpen
  \bibfield  {author} {\bibinfo {author} {\bibfnamefont {I.}~\bibnamefont
  {Ciufolini}},\ }\href {\doibase 10.1088/0264-9381/17/12/309} {\bibfield
  {journal} {\bibinfo  {journal} {Class. Quant. Grav.}\ }\textbf {\bibinfo
  {volume} {17}},\ \bibinfo {pages} {2369} (\bibinfo {year}
  {2000})}\BibitemShut {NoStop}%
\bibitem [{\citenamefont {Abbott}\ \emph
  {et~al.}(2017{\natexlab{b}})\citenamefont {Abbott} \emph
  {et~al.}}]{Abbott:2017GW170814}%
  \BibitemOpen
  \bibfield  {author} {\bibinfo {author} {\bibfnamefont {B.~P.}\ \bibnamefont
  {Abbott}} \emph {et~al.} (\bibinfo {collaboration} {LIGO Scientific
  Collaboration and Virgo Collaboration}),\ }\href {\doibase
  10.1103/PhysRevLett.119.141101} {\bibfield  {journal} {\bibinfo  {journal}
  {Phys. Rev. Lett.}\ }\textbf {\bibinfo {volume} {119}},\ \bibinfo {pages}
  {141101} (\bibinfo {year} {2017}{\natexlab{b}})},\ \Eprint
  {http://arxiv.org/abs/1709.09660} {arXiv:1709.09660} \BibitemShut {NoStop}%
\bibitem [{\citenamefont {Takeda}\ \emph {et~al.}(2021)\citenamefont {Takeda},
  \citenamefont {Morisaki},\ and\ \citenamefont {Nishizawa}}]{Takeda:2021}%
  \BibitemOpen
  \bibfield  {author} {\bibinfo {author} {\bibfnamefont {H.}~\bibnamefont
  {Takeda}}, \bibinfo {author} {\bibfnamefont {S.}~\bibnamefont {Morisaki}}, \
  and\ \bibinfo {author} {\bibfnamefont {A.}~\bibnamefont {Nishizawa}},\ }\href
  {\doibase 10.1103/PhysRevD.103.064037} {\bibfield  {journal} {\bibinfo
  {journal} {Phys. Rev. D}\ }\textbf {\bibinfo {volume} {103}},\ \bibinfo
  {pages} {064037} (\bibinfo {year} {2021})},\ \Eprint
  {http://arxiv.org/abs/2010.14538} {arXiv:2010.14538} \BibitemShut {NoStop}%
\bibitem [{\citenamefont {Kidder}(1995)}]{Kidder:1995}%
  \BibitemOpen
  \bibfield  {author} {\bibinfo {author} {\bibfnamefont {L.~E.}\ \bibnamefont
  {Kidder}},\ }\href {\doibase 10.1103/PhysRevD.52.821} {\bibfield  {journal}
  {\bibinfo  {journal} {Phys. Rev. D}\ }\textbf {\bibinfo {volume} {52}},\
  \bibinfo {pages} {821} (\bibinfo {year} {1995})},\ \Eprint
  {http://arxiv.org/abs/gr-qc/9506022} {arXiv:gr-qc/9506022} \BibitemShut
  {NoStop}%
\bibitem [{\citenamefont {Blanchet}(2014)}]{Blanchet:2014}%
  \BibitemOpen
  \bibfield  {author} {\bibinfo {author} {\bibfnamefont {L.}~\bibnamefont
  {Blanchet}},\ }\href {\doibase 10.12942/lrr-2014-2} {\bibfield  {journal}
  {\bibinfo  {journal} {Living Rev. Rel.}\ }\textbf {\bibinfo {volume} {17}},\
  \bibinfo {pages} {2} (\bibinfo {year} {2014})},\ \Eprint
  {http://arxiv.org/abs/1310.1528} {arXiv:1310.1528} \BibitemShut {NoStop}%
\bibitem [{\citenamefont {Jacobson}(2007)}]{Jacobson:2007}%
  \BibitemOpen
  \bibfield  {author} {\bibinfo {author} {\bibfnamefont {T.}~\bibnamefont
  {Jacobson}},\ }\href {\doibase 10.22323/1.043.0020} {\bibfield  {journal}
  {\bibinfo  {journal} {PoS}\ }\textbf {\bibinfo {volume} {QG-PH}},\ \bibinfo
  {pages} {020} (\bibinfo {year} {2007})},\ \Eprint
  {http://arxiv.org/abs/0801.1547} {arXiv:0801.1547} \BibitemShut {NoStop}%
\end{thebibliography}%

\end{document}